\documentclass[%
 reprint,
 superscriptaddress,
 amsmath,amssymb,
 aps,
 prx
]{revtex4-2}

\usepackage{braket}
\usepackage{graphicx}
\usepackage{dcolumn}
\usepackage{bm}
\usepackage{siunitx}
\usepackage{xcolor} 
\usepackage{tabularx}
\usepackage{dsfont}					
\usepackage{mathrsfs}				
\usepackage{float}
\usepackage{hyperref}

\newcommand{\tr}[0]{\mathbf{tr}}		

\newcommand{\be}[0]{\begin{equation}}	
\newcommand{\ee}[0]{\end{equation}}

\usepackage{hyperref}
\hypersetup{colorlinks, 
	linkcolor={blue!75!black!80!yellow},
	citecolor={blue!75!black!80!yellow}, 
	urlcolor={blue!75!black!80!yellow}
	}

\usepackage[normalem]{ulem}

\begin{document}


\title{Dynamics of photo-induced ferromagnetism in oxides with orbital degeneracy}


\author{Jonathan B. Curtis}
\email[]{joncurtis@ucla.edu}
\affiliation{College of Letters and Science, University of California, Los Angeles, CA 90095, USA}
\affiliation{Department of Physics, Harvard University, Cambridge, MA 02138, USA}
\author{Ankit Disa}
\affiliation{Applied and Engineering Physics, Cornell University, Ithaca, NY 14853, USA}
\affiliation{Max Planck Institute for the Structure and Dynamics of Matter, Hamburg, DE}
\author{Michael Fechner}
\affiliation{Max Planck Institute for the Structure and Dynamics of Matter, Hamburg, DE}
\author{Andrea Cavalleri}
\affiliation{Max Planck Institute for the Structure and Dynamics of Matter, Hamburg, DE}
\affiliation{Clarendon Laboratory, Department of Physics, Oxford University, Oxford, UK}
\author{Prineha Narang}
\email{prineha@ucla.edu}
\affiliation{College of Letters and Science, University of California, Los Angeles, CA 90095, USA}

\date{\today}

\begin{abstract}
By using intense coherent electromagnetic radiation, it may be possible to manipulate the properties of quantum materials very quickly, or even induce new and potentially useful phases that are absent in equilibrium.
For instance, ultrafast control of magnetic dynamics is crucial for a number of proposed spintronic devices and can also shed light on the possible dynamics of correlated phases out of equilibrium.
Inspired by recent experiments on spin-orbital ferromagnet YTiO$_3$ we consider the nonequilibrium dynamics of Heisenberg ferromagnetic insulator with low-lying orbital excitations. 
We model the dynamics of the magnon excitations in this system following an optical pulse which resonantly excites infrared-active phonon modes. 
As the phonons ring down they can dynamically couple the orbitals with the low-lying magnons, leading to a dramatically modified effective bath for the magnons.
We show this transient coupling can lead to a dynamical acceleration of the magnetization dynamics, which is otherwise bottlenecked by small anisotropy.
Exploring the parameter space more we find that the magnon dynamics can also even completely reverse, leading to a negative relaxation rate when the pump is blue-detuned with respect to the orbital bath resonance.
We therefore show that by using specially targeted optical pulses, one can exert a much greater degree of control over the magnetization dynamics, allowing one to optically steer magnetic order in this system.
We conclude by discussing interesting parallels between the magnetization dynamics we find here and recent experiments on photo-induced superconductivity, where it is similarly observed that depending on the initial pump frequency, an apparent metastable superconducting phase emerges.
\end{abstract}

\maketitle

\section{Introduction}
The idea of using strong optical fields to gain control over phases of quantum matter is a tantalizing one. 
Optical control of ferroelectric~\cite{Nova.2019,Li.2019ilv}, structural~\cite{Wall.2018,Sood.2021,Cavalleri.2001,Juraschek.2020,McLeod.2020,Tobey.2008}, superconducting~\cite{Mitrano.2016,Rajasekaran.2016,Buzzi.2021,Budden.2021,Buzzi.20219sr,Cremin.2019,Liu.2020tt}, charge density wave~\cite{Kogar.2020,Lee.2019,Rohwer.2011}, and magnetic orders~\cite{Disa.2021,Afanasiev.2021,Kimel.2009,Liu.2019,Seifert.2019,Gu.2018,Lovinger.2020,Lovinger.20201p,Ron.2020,Torre.2022,Afanasiev.2021b,Seifert.2022,Maehrlein.2018,Kimel.2002,Titov.2021,Disa.2020} have all been proposed theoretically or demonstrated experimentally.
Essentially, by inducing strongly nonequilibrium scenarios, one can explore an enlarged nonequilibrium phase diagram which may allow for the access of novel phenomena and functionalities.

One way to realize such strongly nonequilibrium scenarios is by resonantly driving the system, inducing coherent oscillations in the Hamiltonian, thereby breaking time-translation symmetry and driving the system away from the thermal regime.
These coherent oscillations can potentially excite parametric resonances~\cite{Michael.2020,Knap.2016,Babadi.2017,Dolgirev.2021ouj,Bukov.2015}, induce novel topology~\cite{Oka.2009,Kitagawa.2010,Lindner.2011,Seetharam.2015,Tonielli.2020,Vogl.2021,Nuske.2020}, produce tunable interactions~\cite{Shan.2021,Chaudhary.2020,Mikhaylovskiy.2020,Seifert.2022}, generate non-thermal correlations~\cite{Dolgirev.2020y2b}, and even lead to effective cooling mechanisms~
\cite{Nava.2018,Fabrizio.2018,Werner.2019}.
In particular, the dynamics of correlated Mott insulators hosting orbital degrees of freedom~\cite{Kugel.1982} is known to be quite rich, exhibiting spin-orbital separation~\cite{Muller.2021ic,Wohlfeld.2011}, tunable exchange~\cite{Barbeau.2019,Liu.2018}, and hidden phases~\cite{Li.2018}
Similarly, ferromagnets with high levels of spin-rotation symmetry can exhibit many exotic phenomena away from equilibrium~\cite{Rodriguez-Nieva.2022,Rodriguez-Nieva.2022b,Bhattacharyya.2020}.

The interplay between spin and orbital fluctuations can lead to dramatic effects including magnon softening~\cite{Singh.2010}, entangled spin-orbital phases~\cite{Khaliullin.2000,Khaliullin.2002,Khaliullin.2003,Oles.2005,Oles.2006,Brzezicki.2015,Brink.1999,Khaliullin.2001}, and pronounced magnetic fluctuations~\cite{Feiner.1997,Mochizuki.2001} and subsequent phase transition~\cite{Pal.2018,Wang.2021,Katsufuji.1997}. 
Inspired by recent experiments on ferromagnetic Mott-insulator YTiO$_3$ (YTO)~\cite{Disa.2021}, we examine the nonequilibrium dynamics of magnons in a model quasi-degenerate orbital system driven out of equilibrium by coherently oscillating optical phonons.

In YTO the $3d^1$ conduction band is formed from the titanium $t_{2g}$ shell which naively has a three-fold degeneracy enforced by a cubic lattice symmetry~\cite{Khaliullin.2003}.
However, in YTO and many other compounds, this cubic symmetry is broken at low temperatures by a GdFeO$_3$-type structural distortion, which then lifts the resulting orbital degeneracy~\cite{Pavarini.2004,Pavarini.2005,Zhang.2020mor,Mochizuki.2004,Solovyev.2006,Solovyev.2009,Mizokawa.1999}.
In this case the system has non-degenerate, but potentially low-lying orbital excitations~\cite{Ishihara.2002,Ishihara.2004,Ulrich.2008,Ulrich.2009,Brink.1999,Krivenko.2012}.
In equilibrium settings, the lifted orbital degeneracy leads to a decoupling between spin and orbital excitations and for most magnetic purposes their cross-coupling can be ignored.

Can these orbital excitations, which are essentially absent from equilibrium processes, significantly modify the out-of-equilibrium dynamics?
This question is not purely academic; the ability to control magnetic order on ultrafast timescales~\cite{Torre.2022,Lovinger.2020,Mikhaylovskiy.2020,Kimel.2009} is crucial for many spintronic technologies, and may also help design better ferromagnets which can operate at higher temperatures more efficiently.
We answer this question in the affirmative, provided the orbitals are relatively low-lying and may come close in energy to relevant optically driven degrees of freedom, such as infrared-active phonons, which can reside in the 1-20 terahertz regime. 
In this case, by judiciously choosing the parameters of the optical driving applied, one can speed-up, slow-down, and even reverse the magnetization dynamics.
This then paves the way for novel control routes in quasi-degenerate spin-orbital systems~\cite{Disa.2021,Maimone.2018} as well as potentially other systems~\cite{Afanasiev.2021,Disa.2020,Mikhaylovskiy.2020}.

The remainder of this paper is structured as follows.
In Sec.~\ref{sec:model} we outline the model system considered, and motivate various parameter choices.
Then, in Sec.~\ref{sec:bath} we show how in equilibrium the orbitals essentially serve to provide a bath for angular momentum for the magnons.
In Sec.~\ref{sec:eq} we examine the dynamics of this system in equilibrium and estimate the equilibrium relaxation time-scale.
In Sec.~\ref{sec:noneq} we then explore the nonequilibrium dynamics of this system following a simulated impulsive drive of optical phonons, presenting the main results of this work.
Finally, we discuss the implications of our results in Sec.~\ref{sec:discuss}, where we conclude by discussing interesting parallels between this system and recent experiments on light-induced superconductivity.
In Appendix~\ref{app:ls-coupling} we show how to map the $t_{2g}$ orbital levels into an effective angular momentum.
In Appendix~\ref{app:keldysh-hp} we present details on the nonequilibrium Keldysh technique as applied to the Holstein-Primakoff spin-wave expansion, and in Appendix~\ref{app:two-time} we show how to reduce these equations to a simple equation of motion for the magnon occupation. 

\section{Model}
\label{sec:model}
Here we introduce a simple model for ferromagnetic spins interacting with a quasi-degenerate orbital bath.
Though this is inspired by YTiO$_3$ (YTO), we emphasize we are considering a more abstract model, and we expect our results to be relevant to other high-symmetry ferromagnetic insulators with low-lying orbital excitations.
In particular, we consider a single electron occupying a low-lying $t_{2g}$ orbital manifold.
In the cubic limit there is a large threefold degeneracy which can lead to pronounced orbital fluctuations. 
In reality this cubic degeneracy is lifted by the GdFeO$_3$ structural distortion which renders the crystal structure orthorhombic and induces a finite crystal field splitting $\Delta$ between the lowest and next-lowest orbitals on each Ti site. 
This leads to a model where each site has an orbital pseudospin-1/2 $\hat{\bm\tau}_j$ in addition to the actual electron spin $\hat{\mathbf{S}}_j$ on each site. 

\begin{figure}
    \centering
    \includegraphics[width=\linewidth]{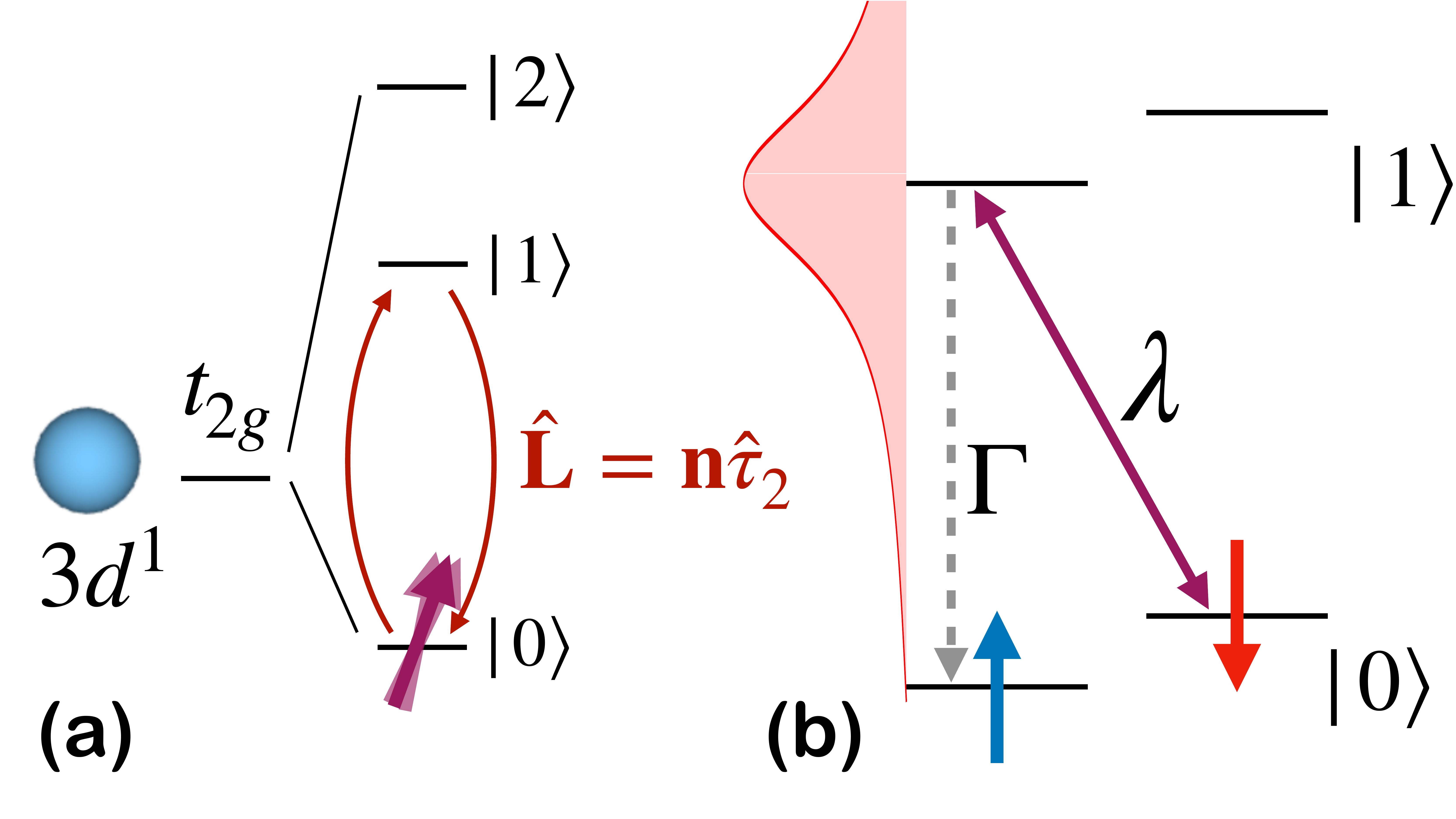}
    \caption{(a) Splitting of Ti $t_{2g}$ shell into non-degenerate levels by the GdFeO$_3$ distortion, which are then occupied by a single electron giving $S = 1/2$.
    Interorbital coherences lead to angular momentum $\mathbf{L}$, which is largely quenched in equilibrium.
    (b) Focusing on the lowest two-levels $|0\rangle$ and $|1\rangle$ we find spin-flip $T_1$ processes in the orbital ground-state obtained from virtual orbital transitions.
    Spin-orbit coupling $\lambda$ can lead to a simultaneous orbital excitation along with a spin-flip.
    This is then followed by a spin-independent orbital decay with rate $\Gamma$, shown in the level diagram.
    Ultimately, the decay rate is governed by the spectral overlap of the orbital bath (shown on the left schematically) with the spin-transition, which is small leading to a long-lifetime. }
    \label{fig:level-diagram}
\end{figure}

We consider a three-dimensional isotropic ferromagnetic Heisenberg model along side a local orbital degree of freedom with Hamiltonian
\begin{multline}
\label{eqn:spin-orbital-H}
\hat{H} = -J \sum_{j,\bm\delta}\mathbf{\hat{S}}_j \cdot\mathbf{\hat{S}}_{j+\bm \delta} - J^z_0\sum_{j,\bm \delta}\hat{S}^z_j\hat{S}^z_{j+\bm\delta}  \\
+ \sum_j \frac{\Delta}{2}\hat{\tau}^3_j + \lambda \mathbf{\hat{L}}_j\cdot \mathbf{\hat{S}}_j .
\end{multline}
The first two terms are the isotropic Heisenberg exchange, with $J\sim 2.75$ meV~\cite{Ulrich.2002} for the case of YTO, and an easy-axis exchange which is chosen to counter the orbital bath-induced Lamb shift, leading to the renormalized spin-wave gap which for the case of YTO was estimated to be 0.02 meV, though the upper bound was qutie a bit larger, of order 0.3 meV~\cite{Ulrich.2002}.
We will consider a modestly-sized renormalized gap of $\Omega_0 = 0.1$ meV in this work.
Here $j$ labels the lattice sites $\mathbf{R}_j$ and ${\bm \delta} = \mathbf{e}_x, \mathbf{e}_y,\mathbf{e}_z$ labels the nearest-neighbors along the three principle axes.

The third term, involving $\hat{\tau}^3_j$, corresponds to the local crystal-field excitation gap.
There is considerable uncertainty about the value of this parameter, with theoretical estimates ranging from nearly zero to over 300 meV.
In Ref.~\cite{Sugai.2006} it was estimated by Raman scattering that $2\Delta\sim$ 50 meV, while Ref.~\cite{Ulrich.2006} found energies closer to $2\Delta \sim$ 235 meV. 
Using resonant inelastic x-ray scattering (RIXS), Ref.~\cite{Ulrich.2009} found evidence for collective orbital excitations with a gap of order 120 meV. 
We will consider $\Delta = 90$ meV here, though more experiments with greater resolution and sensitivity are probably needed in the case of YTO.
The last term describes the atomic $\mathbf{L}\cdot\mathbf{S}$ spin-orbit coupling, which leads to a torque on the spin in the presence of an orbital angular momentum $\mathbf{L}$.

The orbital angular momentum can be obtained by projecting the full three-dimensional $t_{2g}$ angular momentum onto the lowest crystal-field levels, leading to the expression
\begin{equation}
    \mathbf{\hat{L}}_j = \mathbf{n}_j \hat{\tau}^2_{j}.
\end{equation}
The unit vector $\mathbf{n}_j $ is orthogonal to the two participating orbitals, and characterizes the ``soft" axis for orbital angular momentum (see Appendix~\ref{app:ls-coupling}).
The operator $\hat{\tau}^2_j$ characterizes the instantaneous orbital many-body state and in particular $\hat{\tau}^2_j$ is odd under time-reversal (which squares to $+1$ for the $L=1$ orbitals), satisfying the selection rules.
This is coupled to the spin angular momentum by the atomic spin orbit interaction $\lambda$, which in fact need not be small.
Reasonable estimates place $\lambda\sim$ 15 meV for a light $3d$ transition metal such as Ti~\cite{Solovyev.2009}.
Finally, we note that $\mathbf{n}_j$ will in general point in a different direction on each of the four Ti sublattices depending on the local crystal-field environment. 
For more details, we refer to Appendix~\ref{app:keldysh-hp}.
Finally, we will assume that due to, e.g. phonons or orbital interactions the orbital excitation itself obtains a finite $T_1$ linewidth $\Gamma$.
We estimate $\Gamma \sim $ 15 meV as well, although is not known with great certainty and may appear significantly more broad in, e.g a two-orbital spectral function, which may appear in the Raman and RIXS measurements.

We can imagine that in the magnetically ordered phase, each site has a local level scheme as illustrated in Fig.~\ref{fig:level-diagram}(a), where we show the crystal-field splitting of the Ti $t_{2g}$ states and their corresponding spin and orbital angular momentum.
The main idea is that magnetization dynamics is often intrinsically slow due to the bottleneck associated to transfer of spin angular momentum in to a bath, such as the orbital.
One such route is illustrated in Fig.~\ref{fig:level-diagram}(b), which shows how in second order perturbation theory this model can give rise to a finite longitudinal magnetization relaxation rate.
We argue that the phonon dynamics induced by the strong optical pulse can lead to an acceleration of this relaxation time out of equilibrium, leading to the possibility of pump-enhanced magnetization dynamics.  
Note that, unlike Ref.~\cite{Liu.2018} which considered the impact of dynamics on the superexchange interactions, we are more concerned here with the impact on the spin-orbit coupling.

Finally, we comment on the coupling to the drive. 
In the experiment~\cite{Disa.2021}, the pump was performed using a mid-IR pulse which strongly couples to lattice degrees of freedom, rather than e.g. an optical pulse which traverses the Mott gap. 
This pump was tuned to be resonant with various different infrared active phonon modes and used to strongly drive these vibrations. 
Based on {\it ab initio} calculations, we argue that one of the dominant effects of this pump is a strong modulation of the crystal field matrix and in particular, we find that for relevant fluences this may lead to a sizeable change in the eigenvector $\mathbf{n}_j$.
This in turn leads to a dynamical modulation of the orbital angular momentum $\mathbf{L}_j = \mathbf{n}_j(t) \hat{\tau}^2_j$, which will now acquire sidebands at twice the phonon frequency. 

We model this by writing 
\begin{equation}
\mathbf{n}_j(t) \sim \mathbf{n}_j + \delta\mathbf{n}_j Q^2_{\rm IR}(t).
\end{equation}
Here $Q_{\rm IR}$ is the generalized coordinate describing the infrared-active phonon which is directly driven by light (in general there may be different or multiple modes which are excited depending on the frequency and polarization used in the pump and the absoprtion spectrum of the material).
The coupling to $Q^2_{\rm IR}(t)$ is due to the fact that the orbital angular momentum is a Raman active transition whereas the infrared active phonon is polar.
This is in fact very important since this will induce oscillations at {\em twice} the phonon frequency. 
For an $\Omega_{\rm ph} =$ 9 THz optical phonon mode, this leads to sidebands for the spin-orbit coupling at a frequency of $2\Omega_{\rm ph} \sim $80 meV.
This comes close to the orbital resonance in this model at 90 meV.
We will study in particular how the dynamics depends on the drive frequency $\Omega_d$. 
We now proceed to determine the equilibrium structure of this model before proceeding on to compute the nonequilibrium dynamics. 
In particular, we show that the orbitals can act as a bath for angular momentum for the spins even in equilibrium.

Though we do not specifically consider YTO, for certain rough estimates of parameter values and feasibility analysis we have used {\it ab initio} calculations based on YTO.
We estimate interactions between phonon and the crystal field parameters by performing first-principles calculations in the framework of Density Functional Theory (DFT). 
All technical details are listed in Appendix~\ref{app:dft}. 
Our approach is inspired by Ref.~\cite{Zhang.2020mor,Pavarini.2005}, where we first compute the full DFT bandstructure within the local density approximation (LDA). 
Using this we construct localized $t_{2g}$ Wannier-functions using appropriate projectors according to Ref.~\cite{Marzari.1997}. 
To estimate the modulation of the crystal field parameters due to the phonon distortion, we performed frozen phonon computations.  
Therefore, we recalculated the electronic structure and Wannier-functions for crystal structures which have been modulated according eigenvectors of polar eigenmodes ($Q_{\rm IR}$). 
Tthis allows us to estimate changes in the crystal Hamiltonian for distinct polar distortions.

\section{Orbital Bath}
\label{sec:bath}

We now analyze the spin-orbit coupling in equilibrium.
To later accommodate the nonequilibrium calculations, we will implement at the outset the Schwinger-Keldysh formalism for describing this system.
We begin by treating the orbitals in a Gaussian approximation, valid for small orbital excitation amplitudes.
For details we again refer to Appendix~\ref{app:keldysh-hp}, though for a more complete treatment we refer the reader to Ref.~\cite{Kamenev.2011}. 

In the Keldysh formalism we have a doubling of the degrees of freedom, which can be arranged into a ``classical" part $\mathbf{S}_{j,cl}$ characterizing the expectation value, and the ``quantum" part $\mathbf{S}_{j,q}$ which characterizes the fluctuations about the expectation value. 
Applying this formalism to the spin-orbit interaction we find a Keldysh action of 
\begin{equation}
    S_{\rm soc} = - \lambda \sum_{j}\int dt \left[ \mathbf{L}_{jq}(t) \cdot \mathbf{S}_{j,cl}(t) + \mathbf{L}_{jcl}(t) \cdot \mathbf{S}_{j,q}(t) \right].
\end{equation}
This then appears as a contribution to a path integral $Z = \int\mathcal{D}[\mathbf{S},\mathbf{L}] e^{iS}$ which can be used to generate nonequilibrium correlation functions, such as the magnetization $\langle \mathbf{S}_j(t)\rangle$.

We remark to the reader that in this expression, $\mathbf{S}_{j\alpha}(t)$ ($\alpha = cl,q$) should be understood as a stand-in for an appropriate representation of the spin operator in terms of a canonical bosonic or fermionic field.
In this work we will focus on the dynamics in the ordered phase, wherein the operators $\mathbf{S}_j(t)$ can be expanded in terms of the Holstein-Primakoff bosons perturbatively in $1/S$ where $S$ is the spin length.
This is strictly valid only at low-temperatures with $T \ll T_C$ and even then, it suffers from the fact that in YTO $S=\frac12$ is small.
It therefore remains an important problem for future studies to extend this treatment to include the fluctuation regime near $T_C$ by, e.g. an expansion in terms of Schwinger bosons instead, which can better handle the dynamics in the disordered phase. 
Nevertheless, we expect that for low magnon densities, this ought to be at least qualitatively acceptable. 

We proceed by integrating out the local orbital angular momentum, treating it as a bath under a Gaussian approximation.
This bath can be characterized by the correlation functions 
\begin{subequations}
\begin{align}
& \mathbb{\hat{D}}^R(t,t') = -i \langle \mathbf{L}_{j,cl}(t) \mathbf{L}_{j,q}(t')\rangle = \mathbf{n}_j(t) \mathbf{n}_j(t') D^R(t,t') \\
& \mathbb{\hat{D}}^A(t,t') = -i \langle \mathbf{L}_{j,q}(t) \mathbf{L}_{j,cl}(t')\rangle= \mathbf{n}_j(t) \mathbf{n}_j(t') D^A(t,t') \\
& \mathbb{\hat{D}}^K(t,t') = -i \langle \mathbf{L}_{j,cl}(t) \mathbf{L}_{j,cl}(t')\rangle= \mathbf{n}_j(t) \mathbf{n}_j(t') D^K(t,t') .
\end{align}
\end{subequations}
This is in turn expressed in terms of a scalar dynamical response function $D(t,t')$ and the unit vectors $\mathbf{n}_j(t)$, as emphasized in the second equalities, and elaborated on in Appendix~\ref{app:keldysh-hp}.
In equilibrium, these are completely determined given knowledge of the orbital spectral function and the thermal occupation function $\coth(\beta\omega/2)$.

In the Gaussian approximation, we can model the spectral function for $\hat{\tau}^2_j$ as that of a damped harmonic oscillator.
In particular, we assume the orbital angular momentum has linear response equations of motion of  
\begin{subequations}
\begin{align}
    & \frac{d\tau^2}{dt} = \Delta \tau^1 - \Gamma \tau^2 \\
    & \frac{d\tau^1}{dt} = -\Delta \tau^2 - F_{\rm ext}(t).
\end{align}
\end{subequations}
$F_{\rm ext}(t)$ is an external force which acts on the angular momentum $\mathbf{L}\propto \hat{\tau}^2$, which is canonically conjugate to $\hat{\tau}^1$ (the $x$ Pauli matrix whose expectation value corresponds to an interorbital density rather than angular momentum). 
This leads to a spectral function of 
\begin{equation}
    \mathcal{A}(\omega) = -\frac{1}{\pi}\Im \frac{\Delta}{\omega^2 + i\omega \Gamma - \Delta^2} = \frac{1}{\pi} \frac{\omega \Gamma \Delta}{(\omega^2 - \Delta^2)^2 + \Gamma^2 \omega^2}.
\end{equation}
In the limit of $\Gamma \to 0$, this reduces to the spectrum found from the Hamiltonian~\eqref{eqn:spin-orbital-H}.
A realistic estimate for $\Gamma$, based on Raman data~\cite{Sugai.2006} is that $\Gamma \sim 15$ meV, though it seems there is a great amount of uncertainty about this parameter~\footnote{It also seems like there is a good amount of uncertainty about the resonance frequency as well, and whether features are one-orbital, two-orbital, or of other origin all together.}.
From the spectral function, we find the frequency domain Green's functions from Kramers-Kronig relations as 
\begin{subequations}
\begin{align}
& D^R(\omega) =  \frac{\Delta}{\omega^2 + i \Gamma \omega - \Delta^2}  \\
& D^A(\omega) = \frac{\Delta}{\omega^2 - i \Gamma \omega - \Delta^2} \\
& D^K(\omega) = -2\pi i \coth\left(\frac{\beta\omega}{2}\right)\mathcal{A}(\omega) .
\end{align}
\end{subequations}

This then generates an effective action for the spin after integrating out the bath in the Gaussian approximation (for the orbitals) of 
\begin{widetext}
\begin{equation}
    S_{\rm eff} = -\frac{\lambda^2}{2} \int dt \int dt' \sum_j \left( \mathbf{S}_{j,cl}(t') \cdot \mathbf{n}_{j}(t'), \mathbf{S}_{j,q}(t') \cdot \mathbf{n}_{j}(t') \right)\begin{pmatrix}
    0 & D^A(t',t) \\
    D^R(t',t) & D^K(t',t) \\
    \end{pmatrix}\begin{pmatrix}
    \mathbf{n}_j(t) \cdot\mathbf{S}_{j,cl}(t) \\
    \mathbf{n}_j(t) \cdot\mathbf{S}_{j,q}(t) \\
    \end{pmatrix} .
\end{equation}
\end{widetext}
As a final step, we simplify by averaging over the four titanium sublattices.
For a long-wavelength spin-wave, it is reasonable to expect the magnon to only be sensitive to the average of the four titanium sites.
It is worth pointing out that this possibly fails for a short-wavelength magnon, which is localized to the order of one unit cell.
In this case it is possible that the distinct nature of the orbital bath on each site may be important and could be an important source of quantum fluctuations, though we leave this for future work.

\begin{figure}
    \centering
    \includegraphics[width=.7\linewidth]{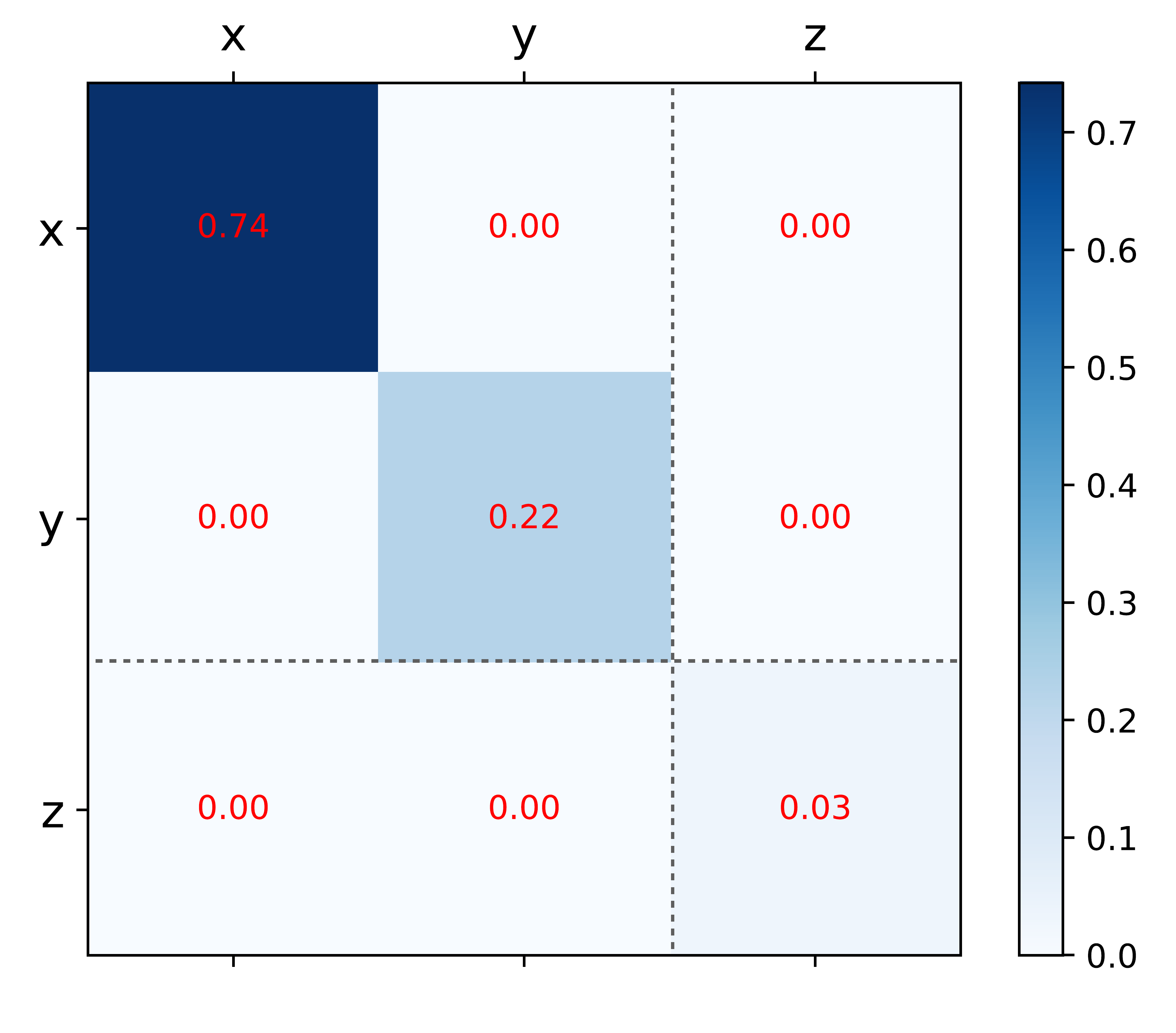}
    \caption{Matrix elements of sublattice averaged anisotropy tensor $\mathbb{N} = \overline{\mathbf{n}\otimes\mathbf{n}}$ in equilibrium.
    We see that the off-diagonal elements are zero, indicating the eigenbasis is aligned with the crystalline axes.
    We also see the eigenvalues are nondegenerate, due to the orthorhombicity of the crystal.
    The matrix is further partitioned into the $ab$-plane ($x,y$) components and the $c$ axis ($z$) components.
    To a good approximation, the matrix projection along the $c$-axis is zero, while the anisotropy in the $ab$-plane is appreciable but not extreme.}
    \label{fig:eq-L-tensor}
\end{figure}

Proceeding on, if we average over the four sites, we generate an effective local action describing the orbital bath of
\begin{widetext}
\begin{equation}
    S_{\rm eff} = -\frac{\lambda^2}{2} \int dt \int dt' \sum_j \left[ \mathbf{S}_{j,cl}(t') \cdot \mathbb{\hat{D}}^A(t',t) \cdot \mathbf{S}_{j,q}(t) + \mathbf{S}_{j,q}(t') \cdot \mathbb{\hat{D}}^R(t',t) \cdot \mathbf{S}_{j,cl}(t) + \mathbf{S}_{j,q}(t') \cdot \mathbb{\hat{D}}^K(t',t) \cdot \mathbf{S}_{j,q}(t) \right].
\end{equation} 
\end{widetext}
This involves the sublattice averaged anisotropy tensor $\mathbb{N}(t',t) = \overline{\mathbf{n}(t')\otimes\mathbf{n}(t)}$ through  
\begin{equation}
    \check{\mathbb{{D}}}(t',t) = \mathbb{N}(t',t) \check{D}(t',t).
\end{equation}
This tensor is presented in Fig.~\ref{fig:eq-L-tensor}, which illustrates the matrix elements along the $x,y,z$ axes (note the $x,y,z$ axes may not necessarily align with the $a,b,c$ axes of the crystal but rather are defined by the orientation of the crystal-field levels).

\section{Equilibrium Spin-Orbit Coupling}
\label{sec:eq}

\begin{figure*}
    \centering
    \includegraphics[width=\linewidth]{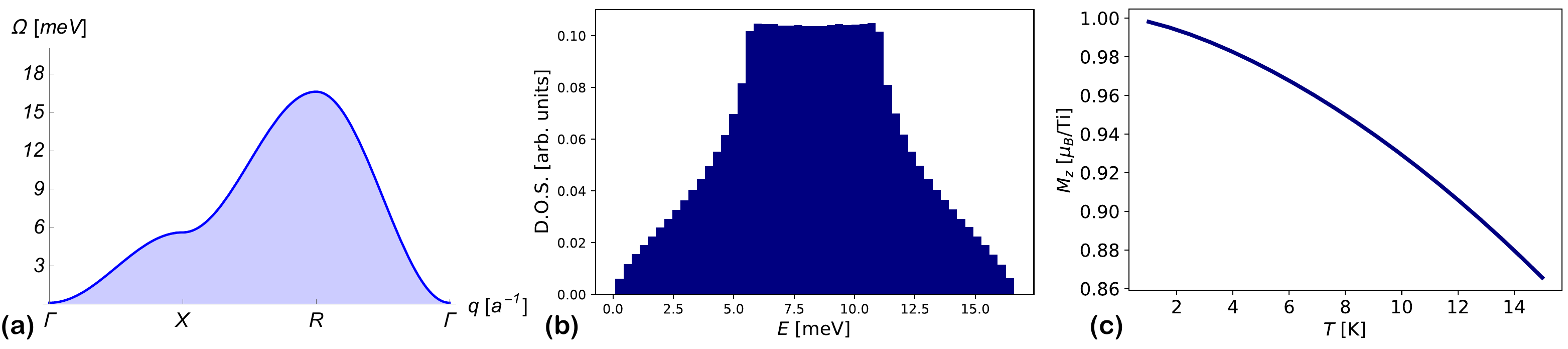}
    \caption{(a) Linear spin-wave dispersion relation along $\Gamma-X-R-\Gamma$ cut in Brillouin zone using idealized cubic model with $J = 2.75$ meV and magnon gap of $\Omega_0 = 0.1$ meV (after renormalizing away the Lamb shift from the orbital bath).
    (b) Magnon density of states (DOS) $\rho(E) = \int_{\bf q} \delta( E - \Omega_{\bf q})$ obtained by Monte Carlo sampling dispersion relation in (a).
    This is used later when we evaluate the integrals of the kinetic equation. 
    (c) Equilibrium occupation of magnons with dispersion in (a), in terms of the magnetic moment $M_z  = 2 \mu_B(S - n_{\rm eq})$.
    We expect the Holstein-Primakoff expansion to underestimate the role of fluctuations since $1/S$ is not small in reality.}
    \label{fig:magnon-dispersion-dos}
\end{figure*}

We now analyze the spin-orbit coupling in equilibrium.
At low temperatures, we can expand around the fully-polarized $|\uparrow \uparrow\uparrow...\rangle$ ground state via the Holstein-Primakoff expansion. 
We then describe magnons in terms of canonical bosons $\hat{b}_j$ via the formal mapping 
\begin{subequations}
    \begin{align}
        & \hat{S}^z_j = S -\hat{b}^\dagger_j \hat{b}_j \\
        & \hat{S}^+_j = \sqrt{2S -\hat{b}^\dagger_j \hat{b}_j}\hat{b}_j \\
        & \hat{S}^-_j = \hat{b}_j^\dagger \sqrt{2S -\hat{b}^\dagger_j \hat{b}_j} .
    \end{align}
\end{subequations}
We then expand in the large-$S$ limit up to order $1/S$ to find the linear spin-wave Hamiltonian. 
The Heisenberg interaction (along with the external field along $\mathbf{e}_z$) gives the standard form, which is diagonalized in momentum space to give 
\begin{equation}
\label{eqn:eq-spin-wave}
    \hat{H}^{(2)}_0 = \sum_{\bf p} \Omega_{\bf p} \hat{b}_{\bf p}^\dagger \hat{b}_{\bf p},
\end{equation}
with dispersion relation (for spin $S=\frac12$)
\begin{equation}
    \Omega_{\bf p} = 6SJ \left[ 1 - \frac13(\cos p_x + \cos p_y + \cos p_z )\right] + 6SJ^z_0 .
\end{equation}
This has a gap set by the easy-plane anisotropy energy $6SJ^z_0$, and has a bandwidth of order $12SJ\sim 18\si{\meV}$. 
At this point, we still need to include the Lamb shift due to orbital fluctuations. 
This dispersion relation is depicted in Fig.~\ref{fig:magnon-dispersion-dos}(a), along with the corresponding single-particle density-of-states (DOS) in Fig.~\ref{fig:magnon-dispersion-dos}(b), computed using Monte Carlo sampling.

We now include the orbital self-energy which can be written in the frequency domain due to the time-translational invariance in equilibrium. 
We expand to order $S$, neglecting the linear term which should vanish when expanding around the ground state. 
We find a Gaussian action for the magnons of 
\begin{equation}
    S^{(2)} = \int_p \overline{b}_p \left[  \check{G}_0^{-1}(p) - \check{\Sigma}(p) \right] b_p  .  
\end{equation}
We expand the orbital bath term to $O(S)$ in Appendix~\ref{app:keldysh-hp} in order to find the magnon self-energy $\check{\Sigma}(p)$.

One can calculate the anisotropy due to the orbital fluctuations and find that due to the orthorhombic nature of YTO, it has three distinct eigenvectors.
Calculating the projections, we find essentially no projection along the $c$ axis, with approximately 75\% and 25\% along the two in-plane directions. 
This has the result of inducing an easy and hard axis for fluctuations, which leads to quantum fluctuations manifested by the anomalous correlation functions of $\langle b_p b_{-p}\rangle$. 
In our simplistic treatment, we neglect these, finding an approximately $U(1)$ system, with dominant eigenvalue of $.5$ and an anisotropy of $.25$ splitting the two principal axes. 

\begin{figure}
    \centering
    \includegraphics[width=\linewidth]{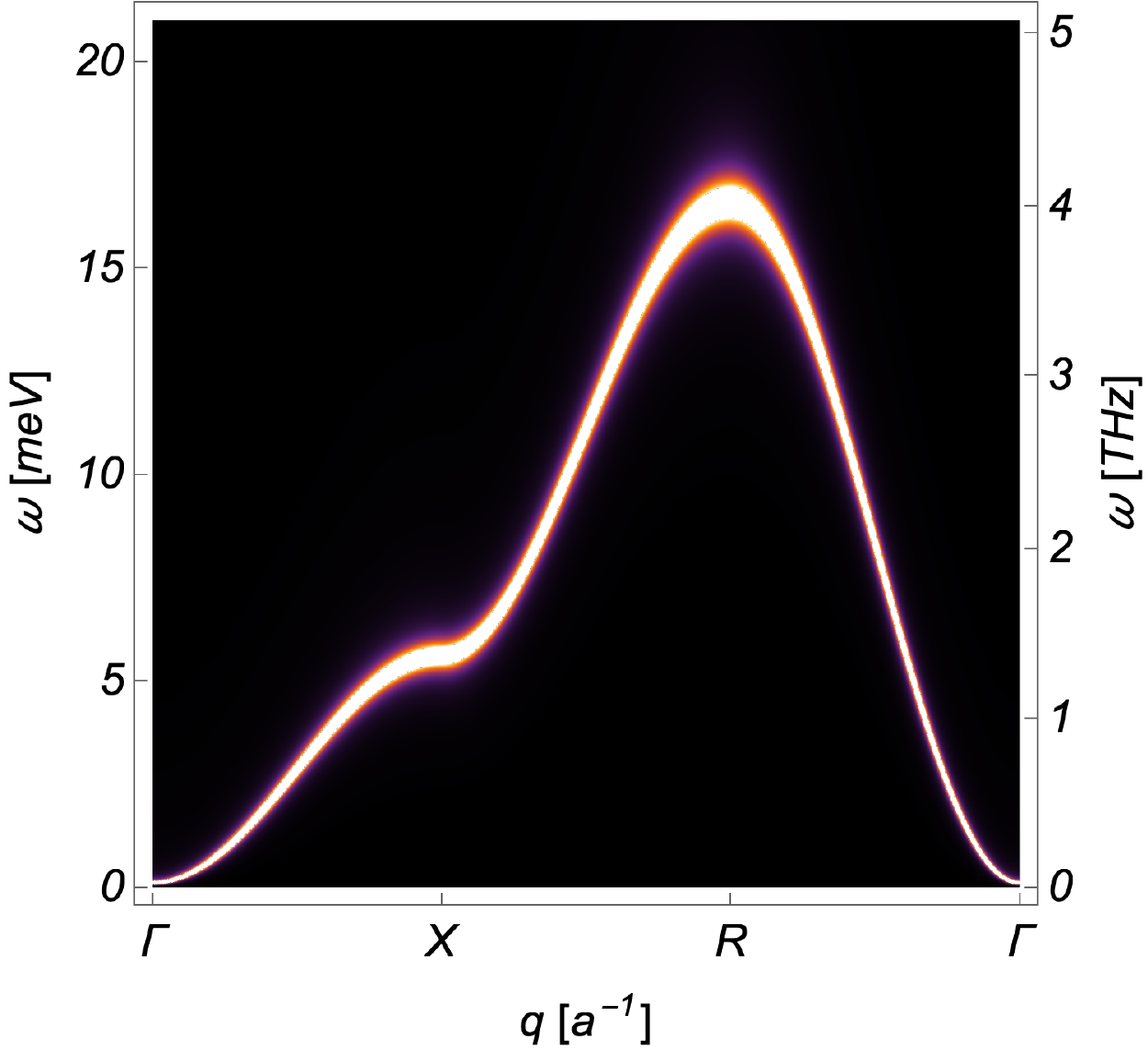}
    \caption{Magnon spectral function $A_{\rm mag}(\omega,\mathbf{p}) = -\frac{1}{\pi}\Im G^R(\omega,\mathbf{p})$ including orbital bath for $N_{+-} = .5$, $\lambda = 15$ meV, $\Delta=90$ meV, and $\Gamma = 10$ meV.
    Plotted along same dispersion contour as Fig.~\ref{fig:magnon-dispersion-dos}. 
    Damping is approximately proportional to frequency, so that the linewidth $\gamma_{\bf p}\sim \Omega_{\bf p}$.
    The Lamb shift due to the orbital bath is renormalized away so that the magnon gap is the physically measured gap of $\Omega_0 = 0.1$ meV. }
    \label{fig:magnon-spectral-func-eq}
\end{figure}

For simplicity, we will neglect the anisotropy so we may obtain a diagonal self-energy, leading to a Green's function of 
\begin{equation}
    G^R(\omega,\mathbf{p}) = \left[ \omega -\Omega_{\bf p} - S \lambda^2 N_{+-} D^R(\omega) \right]^{-1}.
\end{equation}
We have $N_{+-} \sim .48$, which is the isotropic projection of the in-plane components of the anisotropy tensor. 
We plot the magnon spectral function in the $(\omega,\mathbf{p})$ plane in Fig.~\ref{fig:magnon-spectral-func-eq}, however it is worth briefly examining the effects of the orbital bath perturbatively.  

Due to the large separation of scales between the orbital and spin degrees of freedom, we can analyze the corrections to the magnon spectrum perturbatively.
We find a Lamb shift due to the coupling to the reservoir which shifts the magnon band gap (it is essentially a source of single-ion anisotropy of the easy-plane type).
We then find renormalized magnon gap of 
\begin{equation}
    \Omega_0 = 6 SJ_0^z  - \lambda^2 S N_{+-}/\Delta.
\end{equation}
This is used to fix the counterterm $J_0^z$ by matching this to experiment.
It is empirically observed that the gap for magnons is quite small~\cite{Ulrich.2002}, which is in and of itself an interesting fact though we won't dwell on this here.
We also find a finite lifetime is generated for the magnons via their interaction with the bath. 
This has a strong energy dependence and is found to be  
\begin{equation}
    \gamma(E_{\bf p}) = \pi S \lambda^2 N_{+-}\mathcal{A}(E_{\bf p}) = \frac{S \lambda^2 N_{+-}\Gamma E_{\bf p}}{\Delta^3}.
\end{equation}
In particular, the imaginary part scales with $E_{\bf p}$, indicating it is essentially a form of Ohmic Gilbert damping due to the orbital bath. 
Taking estimates for YTO parameters of $\Delta \sim 90$ meV, $\Gamma \sim 15$ meV, and $\lambda\sim 15$ meV, we find a lifetime in ns of
\begin{equation}
    \tau_{\bf p} = 3.6\textrm{ ns}\frac{\textrm{meV}}{E_{\bf p}}.
\end{equation}
We note to the reader that the $\tau_{\mathbf{p}}$ (the lifetime for magnon with momentum $\mathbf{p}$) is completely distinct from $\hat{\tau}^1,\hat{\tau}^2$ which correspond to the orbital operators, and also from $\tau_d$ which corresponds to the lifetime of the phonon ring-down.
These all should occur in separate contexts, but emphasize this distinction here to avoid confusion.
At $T \sim$ 10 K, we have typical magnon energies of $E_{\bf p} \sim 1$ meV and thus we have a typical lifetime for the magnetization relaxation of order 3.6 ns according to this model, though other channels for spin-flip processes may reduce this time according to Matthiessen's rule. 
We now proceed on to study the nonequilibrium dynamics of this system.

\section{Nonequilibrium Dynamics}
\label{sec:noneq}
We now discuss the effect of the strong optical pulse.
Focusing our attention on to the most striking of the three pump frequencies from Ref.~\cite{Disa.2021}, which is the pump at 9 THz, we start by describing how this pump effects the orbital state. 

\begin{figure}
    \centering
    \includegraphics[width=\linewidth]{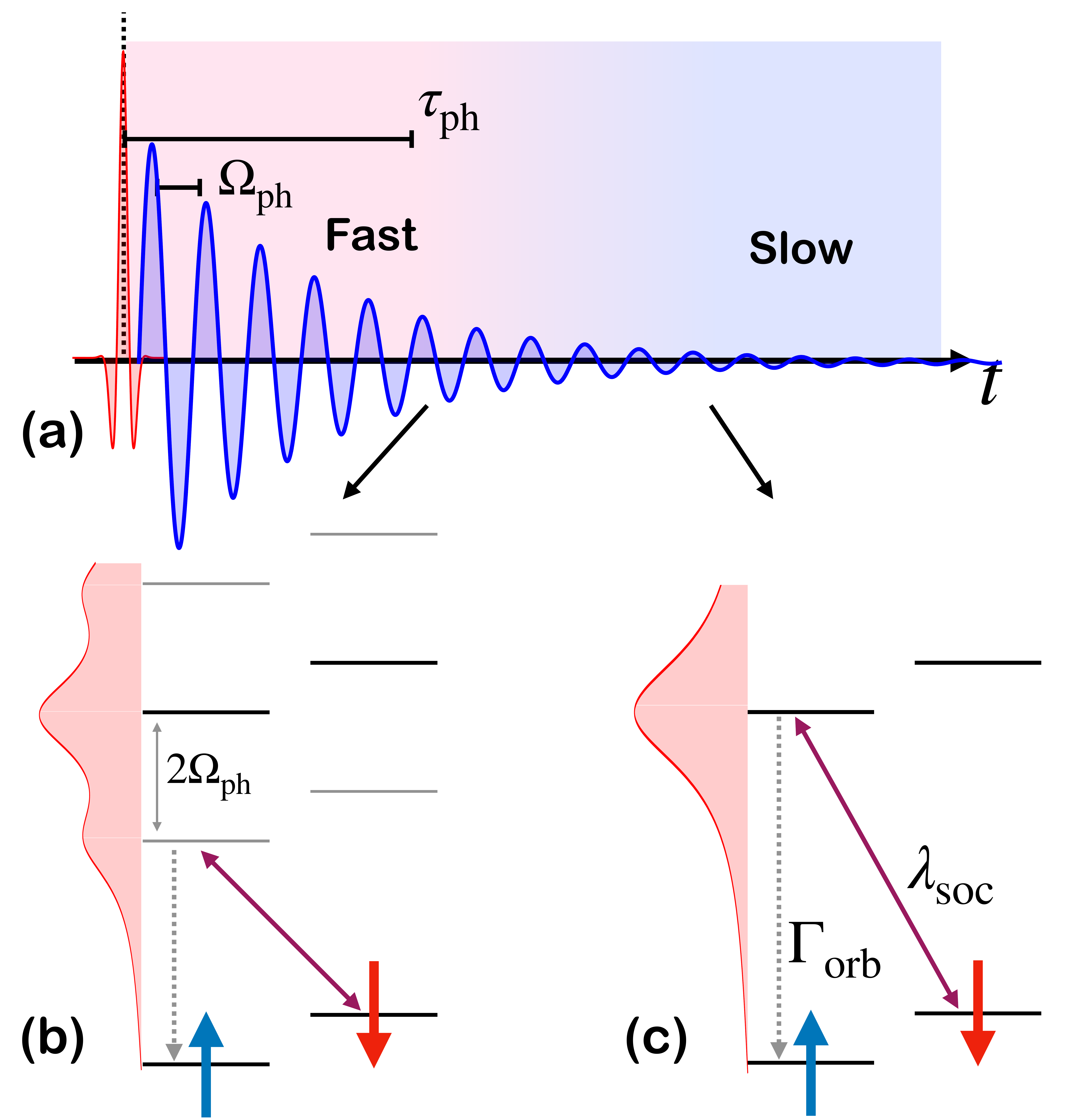}
    \caption{(a) Time scales of pump-induced dynamics. 
    Incident THz pulse (red) resonantly excites a phonon mode which then exhibits coherent oscillations (blue) for time scale $\tau_{\rm ph} \gg 2\pi/\Omega_{\rm ph}$. 
    (b) These dynamics also lead to an acceleration of the spin-orbit mediated magnetization dynamics which leads to faster dynamics during the oscillations due to the appearance of a new channel for spin-flip decay via the phonon-induced sidebands.
    (c) After the oscillations decay, the dynamics returns to the slower time scale present in equilibrium.}
    \label{fig:noneq-levels}
\end{figure}

\subsection{Pump Model}
As per the estimates of the experiment~\cite{Disa.2021}, we consider a terahertz pulse which resonantly drives an IR active phonon mode; in the experiment~\cite{Disa.2021} these were at frequencies of 4 THz, 9 THz, and 17 THz.
Even though the pump itself is quite short, the coherent oscillations it initiates in the phonon mode are estimated to live much longer, with a ring-down time of order of 20-30 ps.
We therefore focus on the magnon dynamics which are induced by these coherent ring-down dynamics rather than the initial pulse which is a quite short duration.
We use a ring-down model of the form 
\begin{equation}
    Q_{IR}(t) = Q_0 e^{-t/\tau_d}\sin(\Omega_d t)\theta(t) ,
\end{equation}
with initial (and maximal) excitation amplitude $Q_0$, central frequency $\Omega_d$, and ring-down time $\tau_d$.

For our purposes, we will assume the pump has two main effects; first, it is assumed to induce a transient change in the spin-exchange $J$ due to a standard spin-phonon coupling mechanism.
The origin of this mechanism is not the main focus of this work, though it may also be interesting.
We simply model this as a coupling between $Q_{\rm IR}$ and $\mathbf{S}_j\cdot\mathbf{S}_{j+\delta}$ of the form 
\begin{equation}
    H_{\rm sp-ph} = \sum_{j,\bm\delta} -\beta \mathbf{\hat{S}}_j\cdot\mathbf{\hat{S}}_{j+\bm\delta}Q_{\rm IR}^2(t).
\end{equation}
This leads to a transient change in $J$ such that we have instantaneous value of $J(t) = J + \beta Q_{\rm IR}^2(t)$.
We focus on the rectified part of this, which leads to a change in the exchange of 
\begin{equation}
    \Delta J(t) = \frac12 \beta Q_0^2 e^{-2t/\tau_d}= \Delta J(0) e^{-2t/\tau_d}.
\end{equation}
We consider two cases \textemdash pump-induced enhancement of $\Delta J(0) = .5 J$ and pump-induced destruction $\Delta J(0) = -.5 J$.

\begin{figure}
    \centering
    \includegraphics[width=\linewidth]{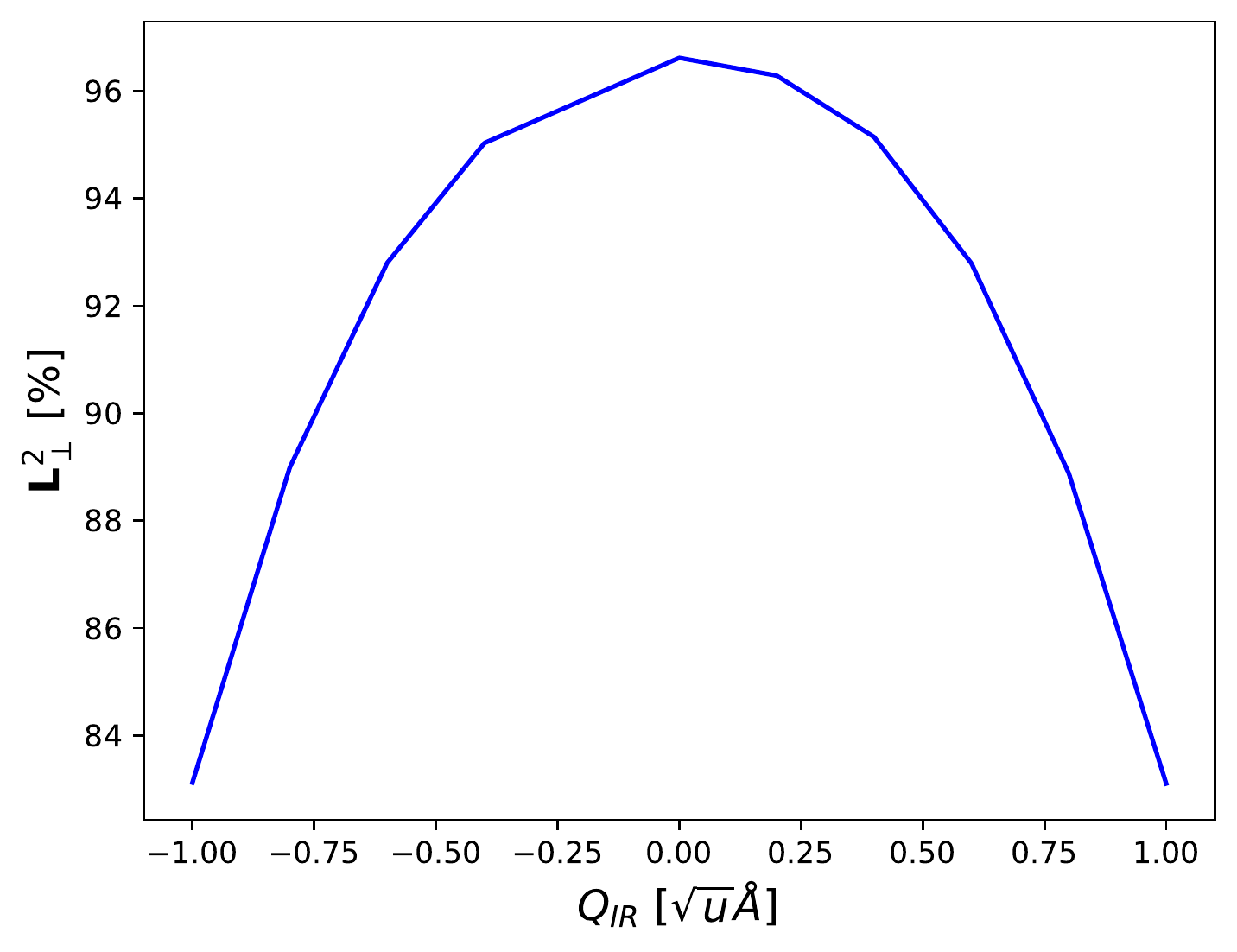}
    \caption{Change in in-plane projection of angular momentum unit vector $\mathbf{n}_j$ for 9 THz $B_{2u}$ polarized phonon mode in YTO.
    We expect for realistic fluence, the peak amplitude is of order $1$ in these units, leading to an appreciable change in the eigenvalues of the anisotropy tensor $\mathbb{N}$ which oscillate at frequencies $\pm 2\Omega_d$.
    This motivates the amplitude parameter $A_d$ of order $A_d \sim \sqrt{.1} = .3$. 
    }
    \label{fig:changes-L}
\end{figure}

In addition to this, we also have argued that the pump induces substantial changes to the excited crystal-field eigenvector, which in turn leads to a dynamical modulation of the spin-orbit coupling between the magnons and orbital bath.
This is motivated by Fig.~\ref{fig:changes-L}, which shows how the orbital angular momentum associated to the first-excited crystal-field transition changes with $Q_{\rm IR}$ in the case of YTO. 

We see there is a quadratic coupling between the phonon mode $Q_{IR}^2$ and the crystal field eigenvector $\mathbf{n}_j$ such that we write 
\begin{equation}
    \mathbf{n}_j \sim \mathbf{n}_j + \delta \mathbf{n} Q^2_{\rm IR}(t).
\end{equation}
This also has a rectified part, which may lead to an interesting pump-induced renormalization of the magnetic anisotropy~\cite{Afanasiev.2021b,Seifert.2022}, however here we will focus instead on the dynamic harmonics, which can dramatically change the nature of magnetic relaxation in this system. 
This is modeled as a change in the anisotropy tensor, which is obtained by averaging this over the four Ti sublattices (we refer to Appendices~\ref{app:keldysh-hp} and~\ref{app:two-time}).

We write this as 
\begin{multline}
    \mathbb{N}(t,t') = \mathbb{N}_{\rm eq}\\
    \times \left[ A_0(t) + A_1(t) e^{-i\Omega_d t} + A_{-1}(t) e^{i\Omega_d t}\right]\\
    \times \left[ A_0(t') + A_1^*(t') e^{+i\Omega_d t'} + A_{-1}^*(t') e^{-i\Omega_d t'}\right]
\end{multline}
where roughly, $A_0^2 + |A_1|^2 + |A_{-1}|^2 = 1$ models the rotation of the excited state unit vector without changing the net length, such that the projection onto the $c$-axis changes by $|A_{-1}|^2 + |A_{-1}|^2$.
We assume the phase is not important, and take 
\begin{equation}
    A_1(t) = A_{-1}(t)  = \frac12 A_d e^{-2t/\tau_d},
\end{equation}
in line with the same pump-profile as the one which drives the change in exchange. 

In order to determine the effect of this Floquet-driven coupling to the orbital bath we will work in the approximation that we may still separate the time scales associated to (i) the magnon dynamics, (ii) the transience of the drive, (iii) the orbital dynamics.
It is very interesting, however challenging, to relax this hierarchy and allow for a complete breakdown of separation of time scales. 
This is left open for future works to handle.
Additionally, though this model captures the essential physics, it is still only qualitatively motivated by YTO calculations, and in future work a more detailed calculation of how exactly the changes in crystal-field evolve for each phonon mode would be warranted. 
In particular, it may be the case that different phonon modes are more or less effective at modulating various components of this tensor and may allow for a more selective control over the effects we describe here.

\subsection{Quasiparticle Dynamics}
We now examine the magnon dynamics in the presence of a hypothetical Floquet modulation of the spin-orbit interaction.
As argued in the previous section, this is a reasonable model of the pumped phonon's effect on the spin-orbit coupling.
To simplify matters, we assume that the pump doesn't actually change the orbital correlations or fluctuations, but rather changes the coupling of the magnons to the orbital bath.

By using the Keldysh technique we are able to calculate the real-time dynamical evolution of the magnon correlation functions, as detailed in Appendix~\ref{app:keldysh-hp}. 
The key object of interest in this work is the magnon occupation function, which is encoded in the Keldysh correlation function $G^K_{\bf p}(t,t')$, here taken to be diagonal in momentum space.
From this, we can then obtain the net magnetization as a function of time. 

We further utilize the separation of time-scales between the evolution under the pump profile and the internal frequency scales by taking the Wigner-transform of $G^K$, which encodes the full two-time dependence in terms of a ``center-of-mass" time, which corresponds to the slow evolution, and the frequency, which encodes the rapid oscillations in the relative time difference.
The Wigner transformed Keldysh function is 
\begin{equation}
    G^K_{\bf p}(T;\omega) = \int d\tau G^K_{\bf p}(T + \frac{\tau}{2}, T-\frac{\tau}{2}) e^{i\omega \tau} . 
\end{equation}
From this, we can extract the total magnon density as a function of time as 
\begin{equation}
    n(t) = \frac{1}{2}\int_{\bf p} \left( \int \frac{d\omega}{2\pi} iG^K_{\bf p}(t;\omega) - 1\right) ,
\end{equation}
and the corresponding magnetization is then found to be 
\begin{equation}
    M^z(t) = 2 \mu_B(S - n(t) ).
\end{equation}

By systematically expanding in terms of gradients of the slowly-varying pump profile, we derive in Appendix~\ref{app:two-time} an effective relaxation-time approximation for this, which to the very lowest order reads
\begin{multline}
    \label{eqn:GK-RTA}
\frac{\partial G^K_{\bf p}(T;\omega)}{\partial T } \\
= 2i\Im \Sigma^R(T;\omega)\left[ -iG^K_{\bf p}(T;\omega) + 2\pi \mathcal{A}_{\rm mag}(T;\omega,\mathbf{p}) F_{\rm orb}(\omega) \right] .
\end{multline}
Here $\mathcal{A}_{\rm mag}(T;\omega,\mathbf{p}) = -1/\pi \Im G^R_{\bf p}(T;\omega)$ is the instantaneous magnon spectral function, which depends on time in the instance where the pump changes, e.g. the spin-exchange, as it does in this system. 
We also see the appearance of the orbital occupation function, which we assume remains in equilibrium at temperature $T_{\rm orb}$, such that $F_{\rm orb}(\omega) = \coth \frac{\omega}{2T_{\rm orb}}$. 

We now study the dynamics of this system under the quasiparticle approximation, such that we can replace the frequency dependence by the instantaneous on-shell frequency.
This gives us a simple equation we can solve for the quasiparticle occupation function $f_{\bf p}(T)$ of 
\begin{equation}
\label{eqn:rta}
    \frac{\partial f_{\bf p}(T)}{\partial T} = -\frac{1}{\tau_{\bf p}(T)}\left(f_{\bf p}(T) - f^{({\rm bath})}_{\bf p}(T) \right),
\end{equation}
where $1/\tau_{\bf p}(T)$ is the instantaneous relaxation rate at time $T$, derived from the magnon self-energy, and $f^{({\rm bath})}_{\bf p}(T) $ is the instaneous equilibrium occupation set by the orbital bath occupation function projected onto the magnon spectral density.
For the details, we refer to Appendix~\ref{app:two-time}.

\begin{figure}
    \centering
    \includegraphics[width=\linewidth]{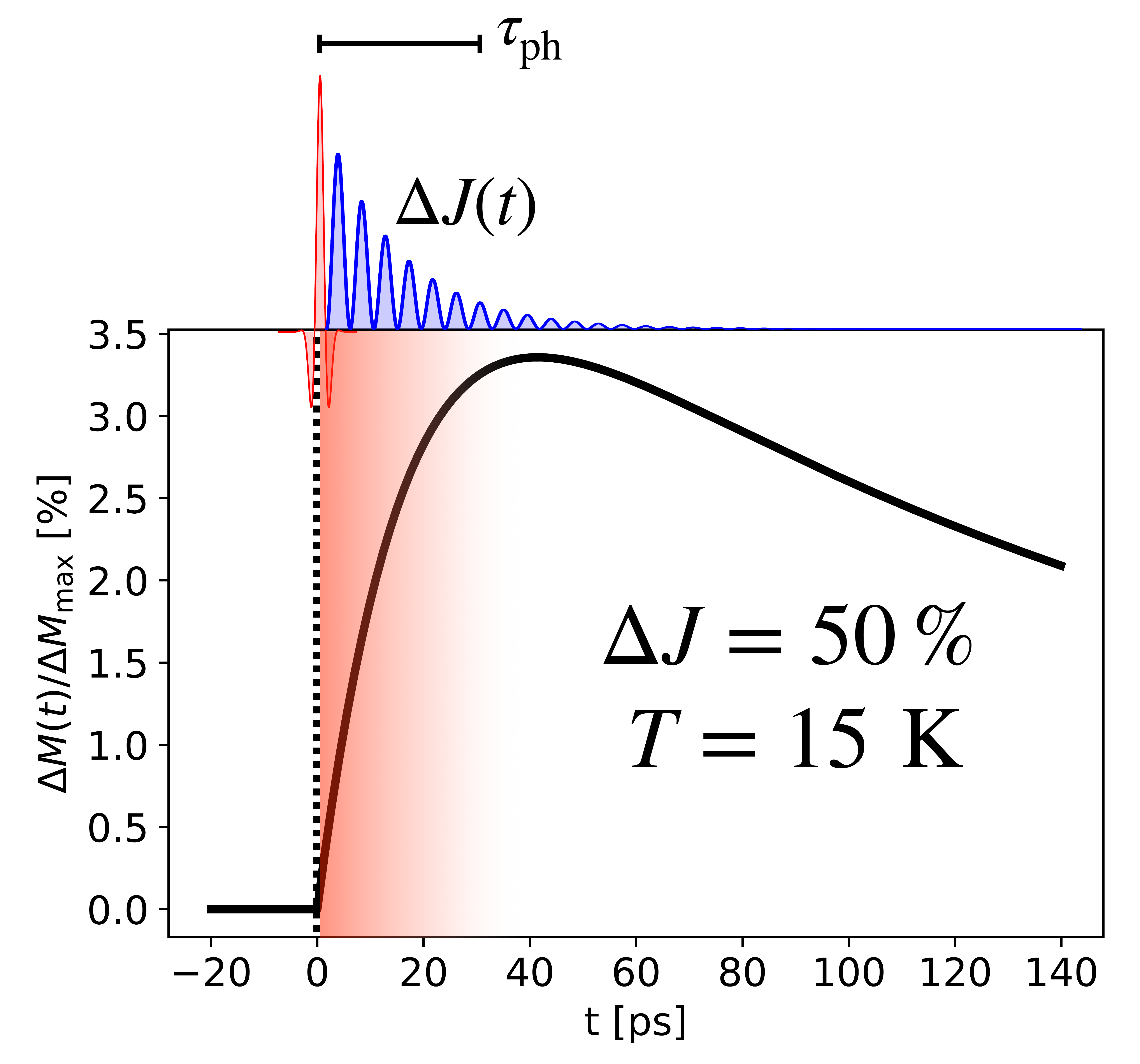}
    \caption{
    Change in magnetization following a pump-induced change in the ferromagnetic exchange due to the rectified spin-phonon coupling, modeled here as a transient $\Delta J(t)\sim Q_{\rm IR}^2(t) $ which follows the impulsive initial pulse, illustrated atop the frame. 
    For a ring-down time of order $\tau_{\rm ph} \sim 30$ ps and an initial change in the exchange of $\Delta J(0)/J_{\rm eq} = 50\%$ we find magnetization dynamics in the plot below, which showns the change in magnetization $\Delta M_z(t)$ in terms a percent of the maximum possible enhancement $\Delta M_{\rm max}$ (corresponding to a complete saturation of the magnetization). 
    Ring-down period is shaded red.
    This is not including the resonant enhancement of the magnetization dynamics. }
    \label{fig:dynamics-no-soc}
\end{figure}

If we only include the change in $J$, and therefore only include the instantaneous change in the spectral function, we see a meager response to the pump.
This is shown in Fig.~\ref{fig:dynamics-no-soc}, which shows the change in instantaneous magnetization following a transient increase in $J$ due to the coherent phonon rind-down, schematically illustrated above the numerical plot. 
We plot the change in magnetization $\Delta M_z(t)$ as a percent relative to the maximum possible change, which would be $2\mu_B(S - n(0))$ so that if the initial moment is $.9\mu_B$ and it increases to $.95\mu_B$ this would by 50\% of the maximum possible increase.

Though the magnetization does generally follow the pump-induced change $\Delta J$, which here was set to $.5 J(0)$, it is a relatively mediocre response since the dynamics are still quite bottlenecked by the long-relaxation time, $\tau_{\bf p}$ which is of order nanoseconds for a thermal magnon, whereas the duration of the pump-induced oscillations are at most 50 ps.

\begin{figure}
    \centering
    \includegraphics[width=\linewidth]{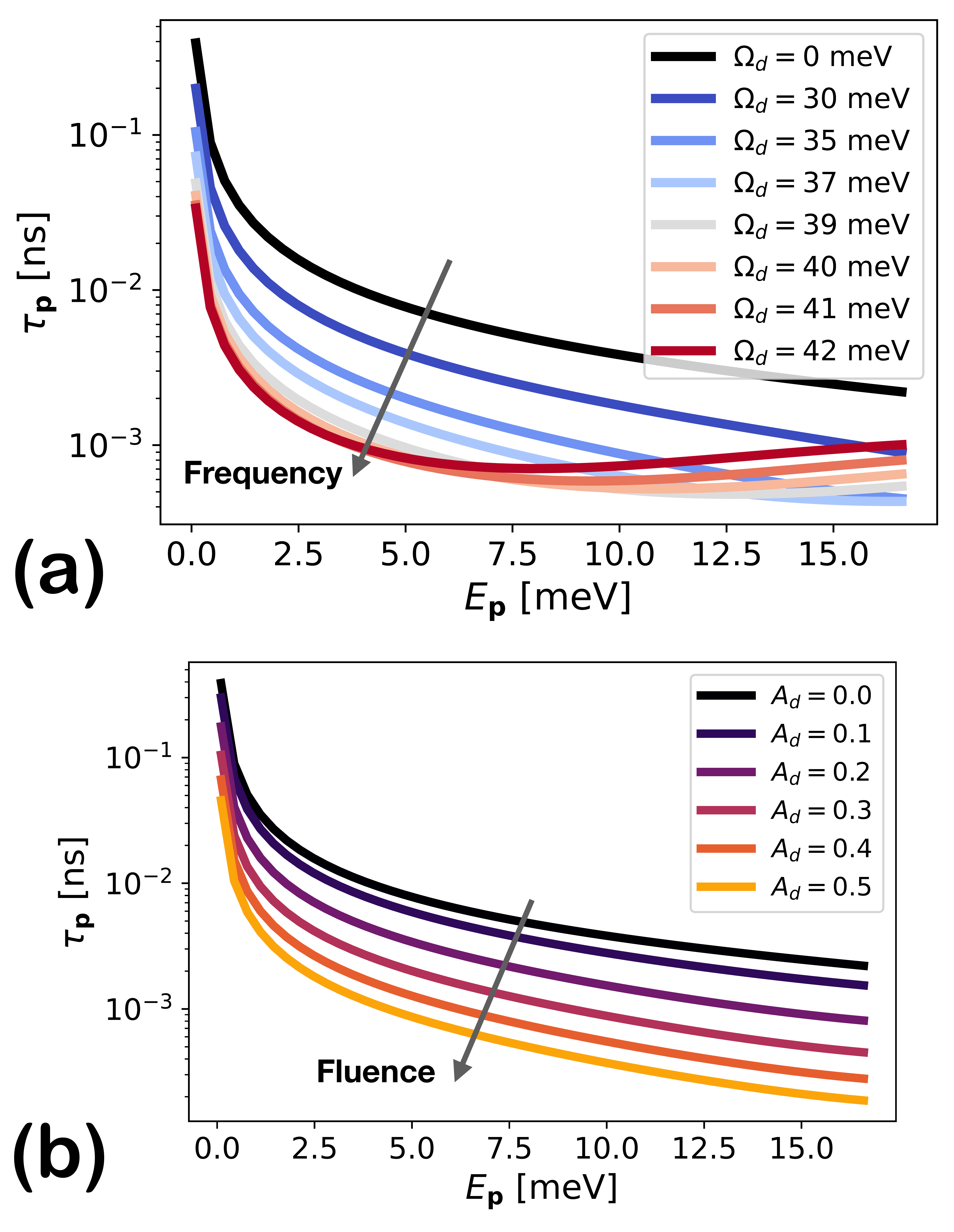}
    \caption{(a) Plot of magnon lifetime $\tau_{\bf p} = 1/\gamma_{\bf p}$ in the presence of the coherent phonon enhancement as a function of magnon kinetic energy $E_{\bf p}$ for different drive frequencies.
    We fix fluence parameter $A_d =.3$ and fix orbital parameters to $\Delta = 90$ meV and $\Gamma = 10$ meV, with $\lambda = 15$ meV. 
    The lifetime is reduced by the magnon appearance of sidebands at $\Delta \pm 2\Omega_d$.
    Near $\Omega_d = 40$ meV this process nears resonance and the decay rate is maximally enhanced by nearly two orderes of magnitude.
    (b) We study for varying drive fluence parameter $A_d$ at fixed $\Omega_d = 35$ meV for the same orbital parameters. 
    The dependence in this model is monotonic, though in a more refinded model we would expect some saturation as $A_d \to 1$. }
    \label{fig:magnon-lifetime}
\end{figure}

However, as we argued before, the nonequilibrium dynamics induced by the pump can potentially have exhibit accelerated timescales, as illustrated in Fig.~\ref{fig:noneq-levels}.
Due to a combination of high-frequency oscillations at $2\Omega_d \sim 80$ meV and low-lying orbital excitations with $\Delta \sim 90$ meV or so, we can find a transient acceleration of the relaxation rate, quantified by $1/\tau_{\bf p} = -2/\pi \Im \Sigma^R(T;\omega)$, making the system essentially relax faster than in equilibrium during the driving period. 
This is confirmed by calculating the effective magnon lifetime in the presence of steady-state coherent oscillations.
In Fig.~\ref{fig:magnon-lifetime} we plot the magnon lifetime $\tau_{\bf p}$ as a function of the magnon kinetic energy $E_{\bf p}$ for different pump frequencies $\omega_d$ and amplitudes $A_d$~\footnote{In fact, $A_d\sim Q_{\rm IR}^2$ is itself scaling linearly with fluence, so that scaling quadratically with $A_d^2$ implies quadratic fluence dependence.}. 

\begin{figure}
    \centering
    \includegraphics[width=\linewidth]{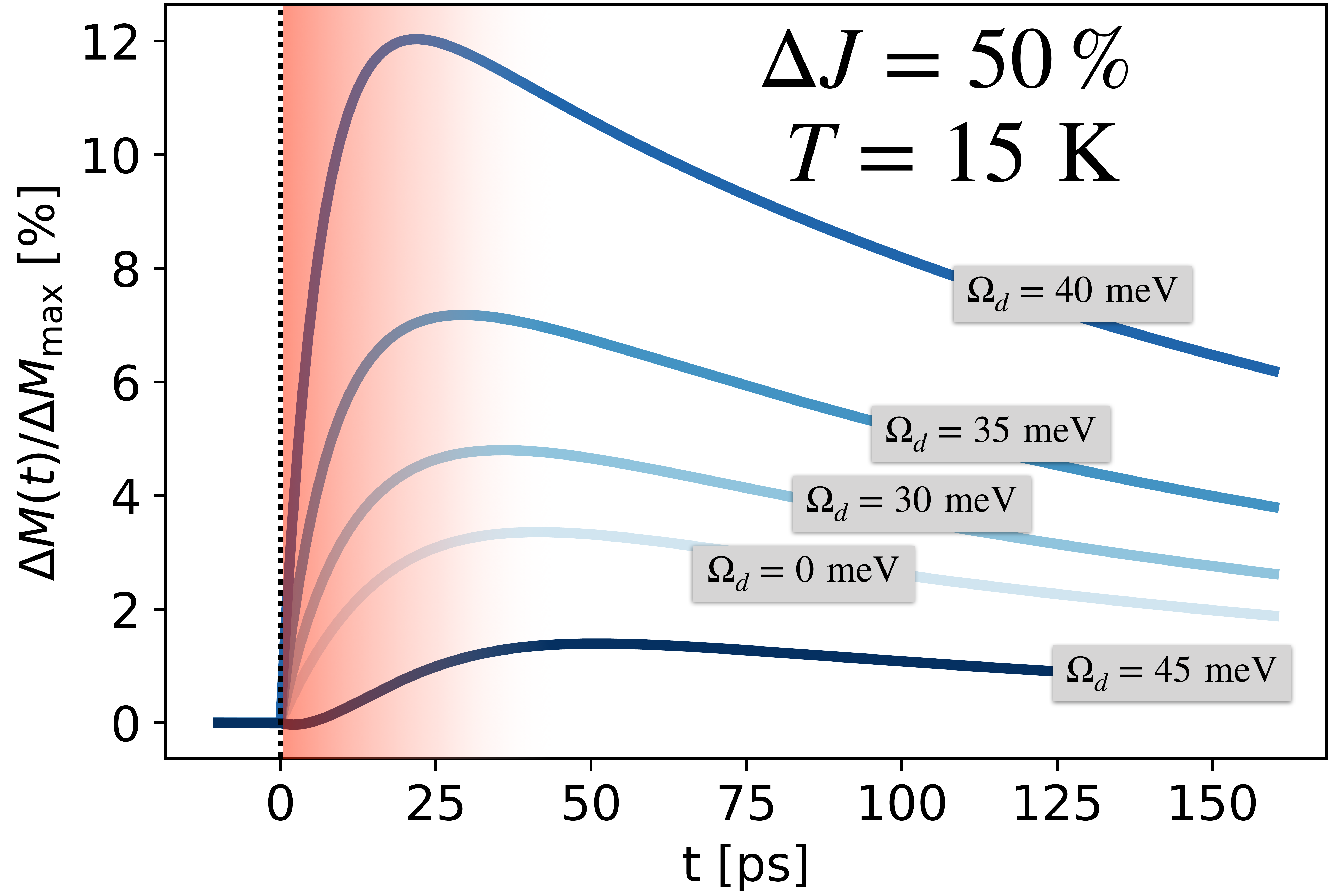}
    \caption{Fractional change in magnon occupation following pump-induced change in exchange $\Delta J$ while also including the enhancement of the relaxation rate due to the phonon ring-down.
    For different pump frequencies (here we only model the pump frequency as changing the spin-flip time) we see a dramatic increase in the maximum change in magnetization upon approaching the resonance condition around $\Omega_d \sim 40$ meV.
    For pump frequencies above this, the effect quickly reverses and by $\Omega_d \sim $ 45 meV we see the dynamics has actually slowed substantially.}
    \label{fig:pump-enhanced-dynamics}
\end{figure}

To see whether the increased relaxation rate has any effect in practice, we carry out the simulations of Eq.~\eqref{eqn:rta} now including both the pump-induced change in $J(t)$ as well as the pump-induced change in relaxation rate. 
This is presented in Fig.~\ref{fig:pump-enhanced-dynamics} which shows the equivalent $\Delta J$ as in Fig.~\ref{fig:dynamics-no-soc} but now including the pump-accelerated relaxation rate for different frequencies $\Omega_d$ at fixed fluence $A_d = .3$. 
We see that when the pump approaches resonance with the orbital excitation, the dynamics greatly accelerates and as a result, the magnetization can grow much more over the same $\sim 30$ ps window of growth time.

\begin{figure*}
    \centering
    \includegraphics[width=\linewidth]{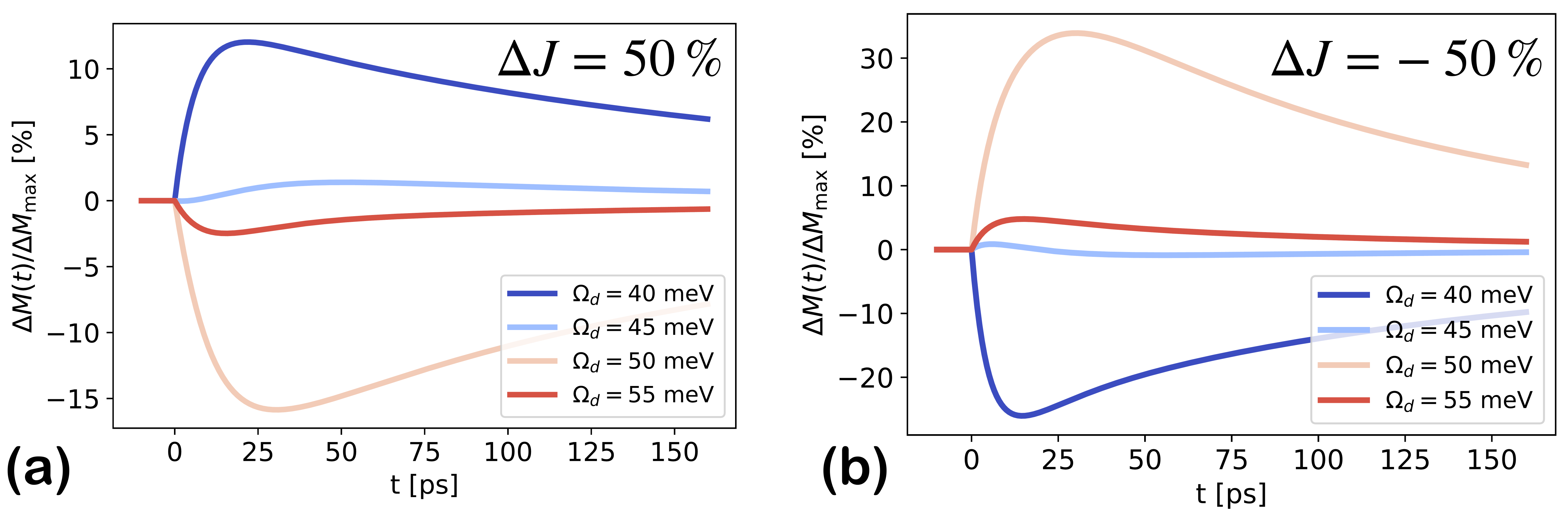}
    \caption{Magnetization dynamics for frequencies below and above resonance.
    (a) For a transient increase in $J$ of 50\% ferromagnetism should increase in equilibrium, however in the presence of a high-frequency drive this can amplify, diminish, or even reverse as the frequency passes through resonance with the bath.
    For $\Omega_d = 40$ meV the relaxation rate reaches near maximal enhancement and the magnetization is most responsive to the pump-induced increase in $J$, while for $\Omega_d = 50$ meV it has already passed to the other side of the resonance.
    The bath now acts to induce ``gain" rather than loss and drives the magnetization opposite to the naive result, quite dramatically.
    (b) If we consider instead a pump-induced reduction in $J$ of =50\% the same features qualitatively persist, with opposite directions. 
    In this case, driving above the resonance leads to a substantial enhancement of magnetization.}
    \label{fig:sign-reversal}
\end{figure*}

Curiously, we see that around $\Omega_d = 45$ meV, the effect seems to completely dissappear, and the resulting magnetization growth is almost completely stunted.
In fact, this is a manifestation of the pump actually passing through the orbital resonance and changing from red-detuning to blue-detuning.
If we continue to increase the drive frequency further, we find that the relaxation rate actually becomes negative\textemdash an effect which is clearly impossible in equilibrium.
This negative relaxation rate essentially indicates that in the rotating frame the orbital bath is population-inverted with respect to the magnon system. 
Therefore, the bath actually acts as a gain medium rather than a retarder.
The resulting dynamics are shown in Fig.~\ref{fig:sign-reversal} where we simulate both an initial increase in exchange, as in Fig.~\ref{fig:pump-enhanced-dynamics}, as well a pump-induced reduction in $J$ of $\Delta J /J(0)= -50\%$. 
We see that the response is most pronounced when $\Omega_d$ is around $\pm 5$ meV detuned from the $\Delta/2 = 45$ meV point. 
We also see that the negative relaxation rate essentially leads to an effectively reversed sign of $\Delta J$, leading to growth in magnon number when it should become less ferromagnetic, and vice versa. 

Therefore, we see that not only can one try to accelerate magnetic dynamics away from equilibrium by modulating the coupling to the orbital bath, but one may even potentially slow the dynamics down (in our example, by tuning $\Omega_d\sim 45$ meV) or reverse them altogether by changing from red- to blue-detuning. 
This is a genuinely nonequilibrium process and may potentially explain the apparent opposite trend between the equilibrium spin-phonon coupling and pump-induced response in YTO in the recent experiment~\cite{Disa.2021}.

\begin{figure*}
    \centering
    \includegraphics[width=\linewidth]{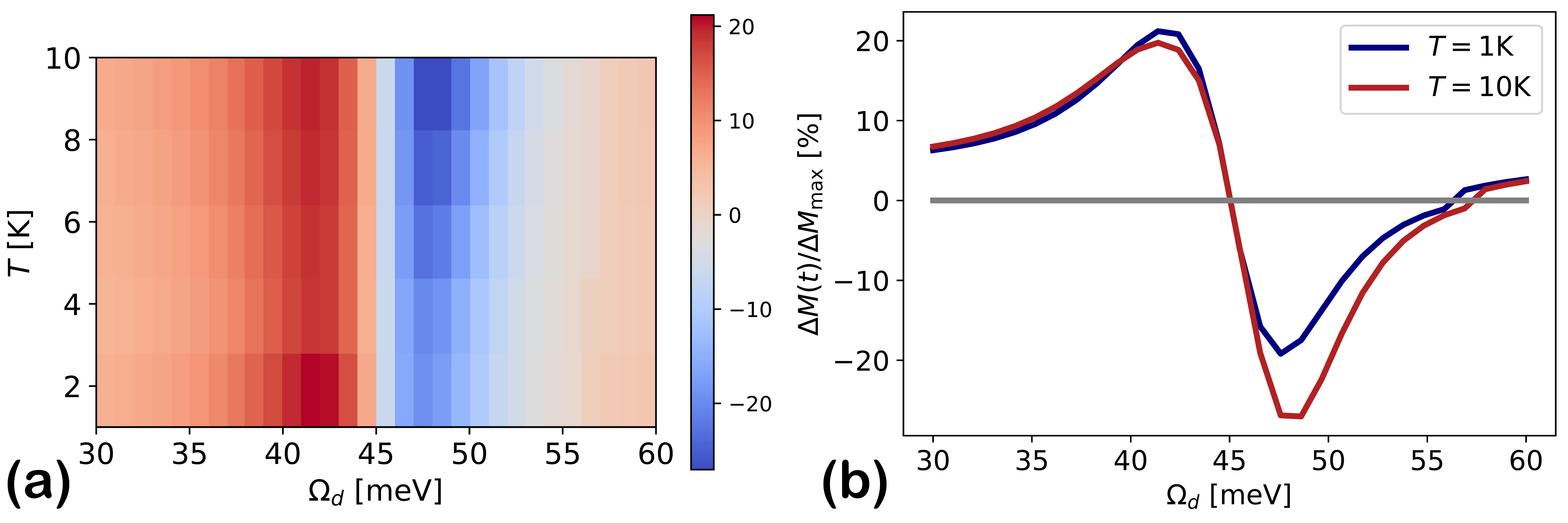}
    \caption{(a) Plot of maximum change in magnetization as a function of initial temperature and pump frequency.
    (a) Color map in $(\Omega_d,T)$ plane. 
    We see that crossing through the resonance at $\Omega_d = \Delta/2$ there is a dramatic change in the sign of the effect, and that the greatest change occurs in this region.
    (b) Line cuts at low temperature ($T = 1$ K) and high temperature ($T = 10$ K).
    We see that the effect is slightly more efficient at increasing the magnetization when temperature is low, while it is more effective at reducing the magnetization at higher temperatures.}
    \label{fig:max-DeltaMz-T-Omega}
\end{figure*}

We can more systematically map this effect out by plotting the most extreme value of the time-traces as a function of $\Omega_d$ and temperature, shown in Fig.~\ref{fig:max-DeltaMz-T-Omega}(a) as a density plot, and in Fig.~\ref{fig:max-DeltaMz-T-Omega}(b) for two line-cuts at fixed temperature $T$.
We see quite clearly that the dynamics are most dramatically affected near the resonance of $2\Omega_d = \Delta$, and upon passing through the resonance the sign of the effect changes.

Thus, we see that going beyond the ``quasistatic" picture and actually considering how the coupling to the orbital bath changes in the presence of nonequilibrium dynamics can lead to striking, and potentially useful changes to the magnetization dynamics.
Crucially, the effect we outline here relies on a relatively low-lying orbital excitation which couples to the spins and also the pumped phonon modes.
If the orbitals are too low-lying then they will exhibit strong fluctuations and cannot be treated as a bath, as we have here.
On the other hand, if they are too high in excitation energy, they cannot be effectively coupled to by phonon oscillations and therefore are cannot realistically participate in the dynamics.
Thus, quasi-degenerate magnetic insulators present a special opportunity for this type of ``bath-control," although as we will discuss next, this type of physics may be able to be extended to more general systems such as antiferromagnets, superconductors, or potentially other correlated phases. 

\section{Discussion}
\label{sec:discuss}

We now summarize our findings.
We considered a simple model for the nonequilibrium dynamics of magnons in a Heisenberg ferromagnetic insulator with low-lying ``quasi-degenerate" orbitals, as may be be realized in orthorhombic titanates $R$TiO$_3$ ($R = $ Y, Sm, Gd), and possibly other compounds.
By using a powerful terahertz pulse to resonantly excited optical phonons we argued that relatively long-lasting nonequilibrium dynamics can be induced by the coherently oscillating phonon modes, which may have lifetimes lasting up to 30 ps. 
These phonon oscillations may lead to transient modifications to the superexchange though, e.g. the rectified part of the spin-phonon coupling $\sim Q_{\rm IR}^2 \mathbf{S}_j\cdot\mathbf{S}_k$; this may then lead to dynamic changes in the magnetic free-energy landscape which can potentially be used to optical drive the magnetization and control the phase diagram.
However, this dynamics is often plagued by a bottleneck due to small spin-orbit coupling which leads to an approximate conservation law for magnetization, leading to slow diffusive dynamics on the relevant time scales. 

This bottleneck can be circumvented in a nonequilibrium setting, as we showed in Sec.~\ref{sec:noneq}.
In particular, in the presence of low-lying orbital excitations, the coupling between magnons and the orbital angular momentum can become unquenched in the presence of phonon dynamics, which may lead to ``stimulated emission" type processes into the orbital bath.
This can in principle lead to a significant acceleration in time scale for the magnetization dynamics, allowing for more effective optical control on relevant time scales.
Furthermore, we found that in principle it is even possible to reverse the nature of the coupling to the bath by changing from red- to blue-detuning with respect to the orbital bath, allowing for an even greater degree of control over the magnetization dynamics.

More generally, our results should be able to be applied to other systems of interest including antiferromagnetic insulators, spin liquids, and other correlated insulators.  
The key component is the ability to induce a dynamical coupling between the degrees of freedom of interest (such as spins) and the bath degrees of freedom.
In addition to controlling the bath decay rates this may also allow to control the bath-induced Lamb shift, which in the case we consider here enters as an effective single-ion anisotropy.
Thus, it may also be possible to control the anisotropy dynamically, as proposed in the recent experiment~\cite{Afanasiev.2021b} and theory~\cite{Seifert.2022}. 
Control over the isotropic superexchange interaction may also be possible through the mechanism we outline here as well as through similar mechanisms~\cite{Mikhaylovskiy.2020,Afanasiev.2021,Barbeau.2019,Gu.2018,Maehrlein.2018}.

Our results may also be relevant to recent experiments on nonequilibrium light-induced superconductivity~\cite{Cavalleri.2018} in fullerides~\cite{Mitrano.2016,Budden.2021,Buzzi.2021}, organics salts~\cite{Buzzi.2020,Buzzi.20219sr}, and cuprates~\cite{Hoegen.2022,Cremin.2019,Liu.2020tt}.
In this case, we argue that there are a number of parallels which make it even more interesting to understand this physics.
Chief among these are the observations of pump-induced signatures of the ordered phase above the equilibrium transition temperature, long-lived resilience of this long-range order, and enhancement of ``coherence" below the ordering temperature.
In the case of superconductors, the effects of pump-induced order are seen most clearly in systems which are strongly coupled and don't exhibit a simple mean-field transition~\cite{Jotzu.2021,Larkin.2005} (e.g. cuprates, fullerides), and this is also the case for the magnetic order in the recent experiment on YTO in Ref.~\cite{Disa.2021}, which appears to exhibit a ``magnetic pseudogap."
It is also possible that a similar equilibrium slowing-down of reaction pathways occurs in these systems, which reside near to a metal=insulator transition~\cite{Imada.1998,Sayyad.2016}.

Although in the current work we don't address the ``pseudogap regime," it should be possible to extend our results to include strong magnetic fluctuations via, e.g. the Schwinger boson technique or various slave-particle mappings, which can be extended to nonequilibrium settings~\cite{Schuckert.2018,Rodriguez-Nieva.2022b,Babadi.2015,Kiselev.2000,Shnirman.2003}. 
It may turn out that the equivalent problem in the superconducting case will actually be more tractable since in this case the theory for a fluctuating superconductor is more amenable to nonequilibrium diagrammatic approaches~\cite{Lemonik.2019,Larkin.2005,Lemonik.2018}.

We also comment that similar ideas have recently been discussed in the context of ``pump-induced sideband cooling" for various solid-state systems by various groups~\cite{Nava.2018,Fabrizio.2018,Werner.2019}.
In particular, it was proposed that recent experiments on light-induced superconductivity~\cite{Mitrano.2016} could be understood by a Floquet sideband cooling utilizing an intermediate bath state provided by an internal excitonic resonance of the fullerene molecules~\cite{Nava.2018}.
This was later extended to the case of a quantum spin-system with a dynamical coupling induced to a complementary bath system~\cite{Fabrizio.2018}.
In this respect, this is very similar to the system we are proposing here, where the orbitals serve as an analogue to the excitonic bath of Ref.~\cite{Nava.2018}.
However, our results should still be present even if the true sideband cooling does not materialize.
In particular, it is likely that both processes will be happening in a true driven system.

To conclude, we have examined the nonequilibrium spin-orbital dynamics in a ferromagnetic insulator and found that away from equilibrium there is a rich variety of dynamical processes which can happen even in a relatively simple quasiparticle description.
Experimentally, this is possibly relevant to various ferromagnetic insulators realized in ferromagnetic rare-earth titanates $R$TiO$_3$, and may be more generally applicable to strongly-correlated spin-orbital systems such as NiPS$_3$~\cite{Afanasiev.2021b,Seifert.2022}, CuSb$_2$O$_6$~\cite{Maimone.2018}, other titanates~\cite{Lovinger.2020,Pavarini.2005,Khaliullin.2000,Gu.2018}, manganites~\cite{Brink.2002,Feiner.1999}, vandates~\cite{Khaliullin.2000,Lovinger.20201p,Fujioka.2010,Khaliullin.2000mud}, and a number of other compounds~\cite{Ruckamp.2005}.
There may also be connections to charge-density wave physics, which can also be manipulated by light~\cite{Kogar.2020,Shi.2019l2}.

We also argued that our results may be analogous to recent experiments on photoinduced superconductivity. 
In future works it will be important to consider extending our results to the strongly fluctuating regime near, and above $T_C$ as well as to incorporate the truly dynamical terms which break time-translational symmetry~\cite{Genske.2015,Seetharam.2015,Babadi.2017}.
In addition, considering systems which do have degenerate orbitals (or exhibit genunie spontaneous orbital ordering) would be of great interest, with many exotic phenomena already known to occur~\cite{Wohlfeld.2011}.
It will also be necessary to develop closer connection to specific materials in order to make contact with current and future experiments.
Experiments using ultrafast x-ray scattering may be able to directly confirm these nonequilibrium dynamics~\cite{Muller.2021ic,Mitrano.2020}, though this is likely to be quite challenging theoretically.

\begin{acknowledgements}
The authors. would like to acknowledge crucial discussions with Pavel Dolgirev, Eugene Demler, Andrey Grankin, Andy Millis, David Hsieh, Mohammad Maghrebi, Benedetta Flebus, Aaron M{\"u}ller, and Zhiyuan Sun.
This work is primarily supported by the Quantum Science Center (QSC), a National Quantum Information Science Research Center of the U.S. Department of Energy (DOE). P.N. acknowledges support as a Moore Inventor Fellow through Grant No. GBMF8048 and gratefully acknowledges support from the Gordon and Betty Moore Foundation as well as support from a Max Planck Sabbatical Award that enabled this collaborative project.

\end{acknowledgements}

\bibliography{references}

\appendix
\section{Orbital Angular Momentum}
\label{app:ls-coupling}
Here we present details of the orbital angular momentum projected onto the $t_{2g}$ states.
The orbital angular momentum of the full $d$ shell is in general characterized by a total $L = 2$ operator.
In the presence of a cubic crystal field splitting, this is further split into two $e_g$ levels with quenched angular momentum, and three $t_{2g}$ levels which in general may have an unquenched effective angular momentum of $L_{\rm eff} = 1$ roughly corresponding to the vector representation of the three orbitals~\cite{Georges.2013}. 

Explicitly, we have the representation of the effective angular momentum operators given in terms of the $t_{2g}$ states $|a\rangle = |yz\rangle, |b\rangle = |zx\rangle, |c\rangle = |xy\rangle $ on site $j$ as 
\begin{equation}
    \hat{L}_j^{l} = -i \epsilon_{lmn} \left( |n;j\rangle \langle m;j | - |m;j\rangle \langle n;j | \right).
\end{equation}
Note that this is odd under time-reversal symmetry and has purely off-diagonal matrix elements written in terms of the Cartesian orbitals.

In the presence of further splitting induced by the GdFeO$_3$ lattice distortion, the orbital angular momentum becomes quenched.
We can then project it onto the lowest two states of the intra-$t_{2g}$ distortion, which we call $|0\rangle$ and $|1\rangle$.
If we express the angular orbital momentum in the basis of the crystal-field matrix we find the relevant operator is 
\begin{equation}
    \mathbf{\hat{L}}_j = \left( -i |1;j\rangle \langle 0;j| - |0;j\rangle\langle 1;j| \right) \mathbf{e}_2 = \mathbf{{e}}_2 \hat{\tau}_j^2,
\end{equation}
where the second equality expresses this in a vector representation in terms of the unit vector which points along the direction $\mathbf{e}_z$ for the $xy$ orbital, and so on in a cyclic way. 
The Pauli matrix $\hat{\tau}_j^2$ is the relevant second-quantized operator resulting from the projection of the full orbital state operator onto the two lowest orbitals.
More generally, we may have the crystal-field matrix evolving with a parameter (such as a phonon coordinate), in which case we express this in terms of the relevant crystal-field wavefunctions as 
\begin{equation}
    \mathbf{\hat{L}}_j = {\mathbf{n}}\hat{\tau}^2_j,
\end{equation}
where the matrix elements of the orbital angular momentum vector are obtained via the Levi-Civita symbol as 
\begin{equation}
    \mathbf{n}_c = \mathbf{e}_a\times\mathbf{e}_b 
\end{equation}
where $\mathbf{e}_a$ is the $a^{\rm th}$ (real, by time-reversal symmetry) eigenvector of the crystal-field.

\section{Keldysh Holstein-Primakoff}
\label{app:keldysh-hp}
We introduce the spin fields on the forward and backward contours, expanded up to $O(S^0)$ as 
\begin{equation}
    \mathbf{S}_{j,\pm} = S \mathbf{e}_3 + \sqrt{S}\left[  \mathbf{e}_- b_{j\pm} + \mathbf{e}_+\overline{b}_{j\pm} \right] - \mathbf{e}_3 \overline{b}_{j\pm}b_{j\pm}
\end{equation}
with 
\begin{equation}
    \mathbf{e}_{\pm}=\frac{1}{\sqrt{2}}(\mathbf{e}_1\pm i \mathbf{e}_2).
\end{equation}
We therefore find the expansion for the classical/quantum fields of
\begin{subequations}
\begin{align}
&     \mathbf{S}_{j,cl} = S\frac{1}{\sqrt{2}}\mathbf{e}_3 + \sqrt{S}\left[  \mathbf{e}_- b_{j cl} + \mathbf{e}_+\overline{b}_{j cl} \right]- \mathbf{e}_3 \frac{1}{\sqrt{2}}(\overline{b}_{jc cl}b_{j cl} + \overline{b}_{j q}b_{j q})\\
&     \mathbf{S}_{j,q} = \sqrt{S}\left[  \mathbf{e}_- b_{j q} + \mathbf{e}_+\overline{b}_{j q} \right] - \mathbf{e}_3 \frac{1}{\sqrt{2}}(\overline{b}_{jcl}b_{j cl} + \overline{b}_{j q}b_{j cl}).
\end{align}
\end{subequations}
We require the products $\mathbf{S}^{\alpha}_{j cl}(t)\mathbf{S}^{\beta}_{j q}(t')$ and $\mathbf{S}^{\alpha}_{j q}(t)\mathbf{S}^{\beta}_{j q}(t')$ up to order $O(S)$.
\begin{widetext}
We find at quadratic order the contributions
\begin{multline}
    \mathbf{S}_{jq}(t)\mathbf{S}_{jcl}(t') = S \left[ \mathbb{T}_{++} \overline{b}_{j q}(t) \overline{b}_{j cl}(t') + \mathbb{T}_{+-}  \overline{b}_{j q}(t) b_{j cl}(t') + \mathbb{T}_{-+}  {b}_{j q}(t) \overline{b}_{j cl}(t') + \mathbb{T}_{--}  {b}_{j q}(t) b_{j cl}(t')  \right]  \\
    - \frac{S}{2}\mathbb{T}_{33} \left( \overline{b}_{j cl}(t) b_{j q}(t) + \overline{b}_{j q}(t) b_{j cl}(t) \right),
\end{multline}
and 
\begin{equation}
    \mathbf{S}_{jq}(t)\mathbf{S}_{jq}(t') = S \left[ \mathbb{T}_{++}  \overline{b}_{j q}(t) \overline{b}_{j q}(t') + \mathbb{T}_{+-}  \overline{b}_{j q}(t) b_{j q}(t') + \mathbb{T}_{-+}  {b}_{j q}(t) \overline{b}_{jq}(t') + \mathbb{T}_{--} {b}_{j q}(t) b_{j q}(t')  \right] .
\end{equation}
Here we have introduced the tensors 
\begin{subequations}
\begin{align}
& \mathbb{T}_{++} = \mathbf{e}_+ \otimes \mathbf{e}_+ \\
& \mathbb{T}_{+-} = \mathbf{e}_+ \otimes \mathbf{e}_- \\
& \mathbb{T}_{-+} = \mathbf{e}_- \otimes \mathbf{e}_+ \\
& \mathbb{T}_{--} = \mathbf{e}_- \otimes \mathbf{e}_- \\
& \mathbb{T}_{33} = \mathbf{e}_3 \otimes \mathbf{e}_3 .
\end{align}
\end{subequations}
We have an effective action due to the orbital bath of 
\begin{multline}
    S_{\rm eff} = -S\frac{\lambda^2}{2}\sum_j \int_{t,t'}\tr \bigg\{ \mathbf{n}_j(t')\otimes \mathbf{n}_j(t)   \cdot \bigg[ D^K(t,t')\left(  \mathbb{T}_{++} \overline{b}_{j q}(t) \overline{b}_{j q}(t') +  \mathbb{T}_{+-} \overline{b}_{j q}(t) {b}_{j q}(t') +  \mathbb{T}_{-+} {b}_{j q}(t) \overline{b}_{j q}(t') +  \mathbb{T}_{--} {b}_{j q}(t) {b}_{j q}(t')\right) \\
    +D^R(t,t')\left(  \mathbb{T}_{++}\overline{b}_{j q}(t) \overline{b}_{j cl}(t') +  \mathbb{T}_{+-}\overline{b}_{j q}(t) {b}_{j cl}(t') + \mathbb{T}_{-+} {b}_{j q}(t) \overline{b}_{j cl}(t') +  \mathbb{T}_{--} {b}_{j q}(t) {b}_{j cl}(t')- \frac12 \mathbb{T}_{33}\left( \overline{b}_{j cl}(t) b_{j q}(t) + \overline{b}_{j q}(t) b_{j cl}(t) \right) \right) \\
    + D^A(t,t')\left(\mathbb{T}_{++} \overline{b}_{j cl}(t) \overline{b}_{j q}(t') +  \mathbb{T}_{+-} \overline{b}_{j cl}(t) {b}_{j q}(t') +  \mathbb{T}_{-+} {b}_{j cl}(t) \overline{b}_{j q}(t') +   \mathbb{T}_{--} {b}_{j cl}(t) {b}_{j q}(t') - \frac12 \mathbb{T}_{33} \left( \overline{b}_{j cl}(t') b_{j q}(t') + \overline{b}_{j q}(t') b_{j cl}(t') \right) \right) \bigg] \bigg\} .
\end{multline}
Here the trace is taken over the spin tensor indices.
\end{widetext}
We will herein replace the sublattice specific angular momentum vectors $\mathbf{n}$ with the sublattice averaged matrix 
\begin{equation}
    \mathbb{N}(t',t) = \overline{ \mathbf{n}_j(t') \otimes \mathbf{n}_j(t) }.
\end{equation}
We also will for the most part throw away the anomalous correlations, assuming they are small, though this may be an interesting direction for the future. 
We then find an effective $U(1)$ symmetry for the magnons, getting 
\begin{widetext}
\begin{multline}
    S_{\rm eff} = -S\frac{\lambda^2}{2}\sum_j \int_{t,t'}\tr \bigg\{ \mathbb{N}(t',t)   \cdot \bigg[ D^K(t,t')\left(   \mathbb{T}_{+-} \overline{b}_{j q}(t) {b}_{j q}(t') +  \mathbb{T}_{-+} {b}_{j q}(t) \overline{b}_{j q}(t') \right) \\
    +D^R(t,t')\left(   \mathbb{T}_{+-}\overline{b}_{j q}(t) {b}_{j cl}(t') + \mathbb{T}_{-+} {b}_{j q}(t) \overline{b}_{j cl}(t')- \frac12 \mathbb{T}_{33}\left( \overline{b}_{j cl}(t) b_{j q}(t) + \overline{b}_{j q}(t) b_{j cl}(t) \right) \right) \\
    + D^A(t,t')\left(\mathbb{T}_{+-} \overline{b}_{j cl}(t) {b}_{j q}(t') +  \mathbb{T}_{-+} {b}_{j cl}(t) \overline{b}_{j q}(t')  - \frac12 \mathbb{T}_{33} \left( \overline{b}_{j cl}(t') b_{j q}(t') + \overline{b}_{j q}(t') b_{j cl}(t') \right) \right) \bigg] \bigg\} .
\end{multline}
At this point we can read out the retarded self-energy
\begin{multline}
\Sigma^R(t,t') = \frac{S \lambda^2}{2}\left[  N_{+-}(t',t)D^R(t,t') + N_{-+}(t,t') D^A(t',t)\right]\\
     - \frac{S \lambda^2}{4}\left(\int dt'' N_{33}(t'',t) D^R(t,t'') \delta(t-t') - \frac12\int dt'' N_{33}(t',t'') D^A(t'',t') \delta(t-t')\right) .
\end{multline}
and the Keldysh self energy as 
\begin{equation}
\Sigma^K(t,t') =c\frac{S \lambda^2}{2}\left[  N_{+-}(t',t)D^K(t,t') + N_{-+}(t,t') D^K(t',t) \right]. 
\end{equation}
\end{widetext}
In the retarded self-energy, the last two terms describe the drive-induced dephasing ($T_2$ process), which only enters when the effective magnon gap due to the orbital fluctuations is time-dependent. 
We will leave this study to future works, and ignore it in this case as we assume the projection of the angular momentum matrix elements is small along the $c$ axis. 

To summarize, once we discard the anomalous terms and the pump-induced dephasing we are left with the magnon self-energies of 
\begin{subequations}
\begin{align}
    & \Sigma^R(t,t') = \frac{S \lambda^2}{2}\left[  N_{+-}(t',t)D^R(t,t') + N_{-+}(t,t') D^A(t',t)\right]\\
    & \Sigma^A(t,t') = \frac{S \lambda^2}{2}\left[  N_{+-}(t',t)D^A(t,t') + N_{-+}(t,t') D^R(t',t)\right]\\
    & \Sigma^K(t,t') = \frac{S \lambda^2}{2}\left[  N_{+-}(t',t)D^K(t,t') + N_{-+}(t,t') D^K(t',t) \right].
\end{align}
\end{subequations}
We are now tasked with using these to solve the equations of motion in the driven case. 
Note that 
\begin{equation}
    N_{+-}(t',t) = \overline{\mathbf{e}_+ \cdot \mathbf{n}(t') \mathbf{n}(t)\cdot\mathbf{e}_-} = N_{-+}(t,t').
\end{equation}

\section{Density Functional Theory Calculations}
\label{app:dft}

We performed our computations with the Vienna {\it ab-initio} simulation package VASP.6.2~\cite{Kresse.1996}. 
For the phonon calculations we used the Phonopy software package~\cite{Togo.2015} and the Wannier90 package for Wannierization~\cite{Marzari.1997}. 
Our computations further utilized pseudopotentials generated within the Projected Augmented Wave (PAW)~\cite{Kresse.1999} method. 
Specifically, we take the following configurations for default potentials: Ti $3p^64s^13d^3$, Y $4s^24p^65s^24d^1$, and O $2s^22p^4$. 
We applied the Local Spin Density Approximation (LsDA) approximation for the exchange-correlation potential, which we augment with the Hubbard $U-J$ parameter to account for the localized nature of the d-states of Ti. 
We use $U = 4$ eV and $J = 0.0$ eV. 
As a numerical setting, we used a $9\times 9\times 7$ Monkhorst~\cite{Monkhorst.1976} generated $k$-point-mesh sampling of the Brillouin zone and a plane-wave energy cutoff of 600 eV. 
We iterate self-consistent calculations until the change in total energy has converged up to $10^{-8}$ eV. 

\section{Two-Time Equations}
\label{app:two-time}
Here we elaborate on the details associated to computing the various non-equilibrium Green's function which enter into the magnon kinetic equation.
We focus on the time-frequency domain transforms, assuming the space and momentum dependencies are trivial. 

To begin with, we invoke the formula for the relation of the Wigner transform of two products.
We consider two correlation functions with known Wigner transforms $A(T_1;\omega_1)$ and $B(T_2;\omega_2)$.
We want the Wigner transform of their convolution, expressed in terms of the two-time functions $A,B$ as 
\begin{equation}
    C(T;\Omega) = \int d\tau e^{i\Omega \tau} \int dt A(T+ \tau/2, t) B(t,T-\tau/2).
\end{equation}
This expression can be found in~\cite{Kamenev.2011} and is formally given as an exponential derivative operation as 
\begin{equation}
    C(T;\Omega) = A(T;\Omega) \exp\left( -\frac{i}{2}\left[ \overleftarrow{\partial_T}\overrightarrow{\partial_\Omega} - \overleftarrow{\partial_\Omega}\overrightarrow{\partial_T} \right]\right)B(T;\Omega).
\end{equation}
This is only useful if one can expand the relevant functions in terms of slowly-varying in both time and frequency, which in turn relies on a separation of scales between the dynamics and frequencies. 

We also use the related formula, relevant for the Wigner transform of the point-wise product, 
\begin{equation}
    D(T;\omega) = \int d\tau e^{i\Omega \tau} A(T+\frac{\tau}{2},T-\frac{\tau}{2})B(T+\frac{\tau}{2},T-\frac{\tau}{2}),
\end{equation}
which yields 
\begin{equation}
    D(T;\Omega) = \int \frac{d\omega}{2\pi} A(T; \Omega - \omega) B(T;\omega) . 
\end{equation}

We now apply this to the Green's functions.
First, we consider the magnon retarded Green's function, which obeys the integral equation 
\begin{equation}
\left(i\partial_t -\Omega_{\bf p} \right) G^R(t,t') - \int dt'' \Sigma^R(t,t'') G^R(t'',t') = \delta(t-t').
\end{equation}
In the absence of the drive, this is solved in the frequency domain, and we obtain the standard result which in particular amounts to a form of Gilbert damping at low frequencies. 

In this work we will still retain the separation between the evolution times, which are of order of 20-2000 ps, and the time-scales of the internal degrees-of-freedom which are from 20-800 fs or so.
This allows us to efficiently employ the equations of motion using the Wigner transformations and the Moyal expansions. 

The lowest order in the Moyal expansion is simply the product. 
We retain expansion up to first order, giving equation of motion for the retarded Green's function 
\begin{widetext}
\begin{equation}
\frac{i}{2}\left[ 1 - \partial_\omega \Sigma^R(T;\omega) \right]\partial_T G^R_{\bf p}(T;\omega) + \left[ \omega - \Omega_{\bf p} - \Sigma^R(T:\omega) + \frac{i}{2}\partial_T\Sigma^R(T;\omega) \partial_\omega\right] G^R_{\bf p}(T;\omega) = \mathds{1}. 
\end{equation}
\end{widetext}
This yields a first order differential equation for the Green's function, though it remains non-local in frequency space due to the the changing self-energy.
When solving, we also supplement with the initial condition that 
\begin{equation}
    G^R_{\bf p}(-\infty.\omega) = \frac{1}{ \omega - \Omega_{\bf p} - \Sigma^R(\omega) } .
\end{equation}

To complete this, we need to express the self-energy as a Wigner transform as well.
We use the product formula to find Wigner-transform (applied to $R,A,K$ self-energies)
\begin{equation}
  \check{\Sigma}(T;\Omega) = \frac{1}{N^{\rm eq}_{+-}}\int\frac{d\omega}{2\pi}N_{+-}(T;\omega)\check{\Sigma}_{\rm eq}(\Omega - \omega) .
\end{equation}
Here $\Sigma_{\rm eq}(\omega)$ is the equilibrium self-energy and depends only on frequency.
$N_{+-}^{\rm eq}$ is the equilibrium angular momentum projection, while $N_{+-}(T;\omega)$ is the Wigner transform of the modulated angular momentum tensor. 
We model the modulation via 
\begin{multline}
     N_{+-}(t,t')=  N_{+-}^{\rm eq} (A_0(t) + A_1(t) e^{-i\Omega_d t} + A_{-1}(t) e^{i\Omega_d t} )\\
     \times (A_0^*(t') + A_1^*(t') e^{i\Omega_d t'} + A_{-1}^*(t') e^{-i\Omega_d t'} ),
\end{multline}
where we have expressed this in terms of a Floquet expansion in the drive-frequency $\Omega_d$, along with slowly varying envelope functions $A_0,A_{\pm 1}$, which vary over times of order $\tau_d \gg \Omega_d^{-1}$. 
\begin{widetext}
This gives, in the slowly varying envelope approximation for $A$'s of 
\begin{multline}
     N_{+-}(t,t')/N_{+-}^{\rm eq} = \left[ |A_0(T)|^2 + A_1(T) A_{-1}^*(T) e^{-2i\Omega_d T} + A_{-1}(T) A_1^*(T) e^{2i\Omega_d T}\right] 2\pi \delta(\omega) \\
     + |A_1(T)|^2 2\pi \delta(\omega - \Omega_d) + |A_{-1}(T)|^2 2\pi \delta(\omega + \Omega_d) \\
     + \left[A_0(T) A_1^*(T) e^{i\Omega_D T} + A_{0}^*(T)A_1(T)  e^{-i\Omega_D T} \right]2\pi \delta(\omega - \Omega_d/2) \\
     + \left[A_0(T) A_{-1}^*(T)  e^{-i\Omega_D T} + A_{0}^*(T)A_{-1}(T)  e^{i\Omega_D T} \right]2\pi \delta(\omega + \Omega_d/2) .
\end{multline}
\end{widetext}
 
This involves a number of terms, including some which couple the slow-dynamics to the fast degrees of freedom. 
These terms involve oscillatory couplings like $e^{i\Omega_dT}$.
While these are important close to parametric resonance, or in the steady-state Floquet system, where the separation of time scales completely disintegrates, or must be treated non-perturbatively, we limit ourselves to the regime where the dynamics are still able to be disentangled.
We therefore only keep in this expansion those terms which don't average out over long times $T$.
This leaves only the terms 
\begin{multline}
    \check{\Sigma}(T;\omega) = |A_0(T)|^2\check{\Sigma}(\omega) \\
    + |A_1(T)|^2 \check{\Sigma}(\omega - \Omega_d) + |A_{-1}(T)|^2 \check{\Sigma}(\omega + \Omega_d) .
\end{multline}

In fact, the object we are interested in is the magnon Keldysh occupation function, whose equal-time value reflects the time-dependence of the total number of magnons.
At the Gaussian level, one can find that the this Green's function is given by 
\begin{equation}
    \mathbb{G}^K = \mathbb{G}^R \circ \Sigma^K \circ \mathbb{G}^A .
\end{equation}
\begin{widetext}
In order to proceed, we manipulate this to obtain an equation of motion of the form 
\begin{equation}
    (\mathbb{G}^R)^{-1}\circ \mathbb{G}^K - \mathbb{G}^K \circ (\mathbb{G}^A)^{-1} = - ( \mathbb{G}^R \circ \Sigma^K - \Sigma^K \circ \mathbb{G}^A ) . 
\end{equation}
We now utilize the fact that this is diagonal in momentum space and take the Wigner transform of this equation.
In general, this will not yield a closed form since the Wigner transform is over convolutions of the functions. 
In the very lowest-order limit of a slowly-varying change in the self-energy, we get 
\begin{equation}
    \left[ i\frac{\partial}{\partial T} - ( \Sigma^R(T;\omega) - \Sigma^A(T;\omega) )\right]\mathbb{G}^K_{\bf p}(T;\omega) = -\left(\mathbb{G}^R_{\bf p}(T;\omega) - \mathbb{G}^A_{\bf p}(T;\omega) \right)\Sigma^K(T;\omega).
\end{equation}
This is formulated in terms of the occupation and spectral functions as 
\begin{equation}
    \frac{\partial \mathbb{G}^K_{\bf p}(T;\omega)}{\partial T}= \left[\Sigma^R(T;\omega) - \Sigma^A(T;\omega)\right]\left( -i\mathbb{G}^K_{\bf p}(T;\omega)+2\pi \mathcal{A}_{\rm mag}(T;\omega,{\bf p})F_{\rm orb}(T;\omega) \right).
\end{equation}
\end{widetext}
This is the simple frequency-dependent relaxation-time approximation. 
We find a relaxation of the instantaneous magnon occupation towards the bath temperature with the relaxation rate given by the bath coupling. 

To conclude, we implement the quasiparticle approximation, which assumes the linewidth of the magnon is much smaller than its central frequency.
In this case we can derive a simple equation solely for the total magnon occupation function
\begin{equation}
    f_{\bf p}(T) = i \int \frac{d\omega}{2\pi} G^K_{\bf p}(T;\omega)
\end{equation}
as 
\begin{equation}
    \partial_t f_{\bf p}(t) = -\frac{1}{\tau_{\bf p}(t)} \left[ f_{\bf p}(t) - f_{\bf p}^{(0)}(t) \right],
\end{equation}
where the instantaneous relaxation rate is given by 
\begin{equation}
   \frac{1}{\tau_{\bf p}(t)}  = - 2 \int \frac{d\omega}{2\pi} \Im \Sigma^R(\omega;t ) \mathcal{A}_{\rm mag}(\omega,\mathbf{p};t),
\end{equation}
and the instantaneous occupation function is 
\begin{equation}
   f_{\bf p}^{(0)}(t) = \frac{\int \frac{d\omega}{2\pi} \Im \Sigma^R(\omega;t ) \mathcal{A}_{\rm mag}(\omega,\mathbf{p};t) F_{\rm orb}(\omega) }{\int \frac{d\omega}{2\pi} \Im \Sigma^R(\omega;t ) \mathcal{A}_{\rm mag}(\omega,\mathbf{p};t)}.
\end{equation}
We approximate the spectral function as 
\begin{equation}
    \mathcal{A}_{\rm mag}(\omega,\mathbf{p};t) = \delta(\omega - \Omega_{\bf p}(t) ),
\end{equation}
since the quasiparticle decay rate is expected to be small.

\end{document}